\documentclass[reprint,
superscriptaddress,
nofootinbib,
aps,
prd
]{revtex4-2}

\usepackage{graphicx}
\usepackage{xcolor}
\usepackage{amsmath}
\usepackage{amssymb,gensymb}
\usepackage{bm}
\usepackage{mleftright}
\usepackage[ISO]{diffcoeff}
\usepackage{natbib}
\usepackage{hyperref}
\usepackage[capitalise]{cleveref}
\usepackage{multirow}

\makeatletter
\input{aas_macros.sty}
\let\jnl@style=\relax
\makeatother

\newcommand{\Msun}{\ensuremath{\mathrm{M}_\odot}}

\begin{document}

\title{Tomography of the gamma-ray sky from cross-correlation with DESI DR2 and unWISE galaxies}
\author{Alex Krolewski}
\affiliation{Waterloo Centre for Astrophysics, University of Waterloo, Waterloo, Ontario N2L 3G1, Canada}
\affiliation{Department of Physics and Astronomy, University of Waterloo, Waterloo, Ontario N2L 3G1, Canada}
\author{Neal Dalal}
\affiliation{Perimeter Institute for Theoretical Physics, 31 Caroline Street N., Waterloo, Ontario, N2L 2Y5, Canada}
\author{Will J.\ Percival}
\affiliation{Waterloo Centre for Astrophysics, University of Waterloo, Waterloo, Ontario N2L 3G1, Canada}
\affiliation{Department of Physics and Astronomy, University of Waterloo, Waterloo, Ontario N2L 3G1, Canada}
\affiliation{Perimeter Institute for Theoretical Physics, 31 Caroline Street N., Waterloo, Ontario, N2L 2Y5, Canada}
\author{Elena Pinetti}  
\affiliation{Center for Computational Astrophysics, Flatiron Institute, New York, NY 10010, USA}
\affiliation{Emmy Noether Fellow, Perimeter Institute for Theoretical Physics, 31 Caroline Street N., Waterloo, Ontario, N2L 2Y5, Canada}
\affiliation{Theoretical Physics Division, Fermi National Accelerator Laboratory, Batavia, IL, USA}
\affiliation{Kavli Institute for Cosmological Physics, University of Chicago, Chicago, IL USA}

\author{J.~Aguilar}
\affiliation{Lawrence Berkeley National Laboratory, 1 Cyclotron Road, Berkeley, CA 94720, USA}

\author{S.~Ahlen}
\affiliation{Department of Physics, Boston University, 590 Commonwealth Avenue, Boston, MA 02215 USA}

\author{S.~BenZvi}
\affiliation{Department of Physics \& Astronomy, University of Rochester, 206 Bausch and Lomb Hall, P.O. Box 270171, Rochester, NY 14627-0171, USA}

\author{D.~Bianchi}
\affiliation{Dipartimento di Fisica ``Aldo Pontremoli'', Universit\`a degli Studi di Milano, Via Celoria 16, I-20133 Milano, Italy}
\affiliation{INAF-Osservatorio Astronomico di Brera, Via Brera 28, 20122 Milano, Italy}

\author{D.~Brooks}
\affiliation{Department of Physics \& Astronomy, University College London, Gower Street, London, WC1E 6BT, UK}

\author{T.~Claybaugh}
\affiliation{Lawrence Berkeley National Laboratory, 1 Cyclotron Road, Berkeley, CA 94720, USA}

\author{A.~Cuceu}
\affiliation{Lawrence Berkeley National Laboratory, 1 Cyclotron Road, Berkeley, CA 94720, USA}

\author{A.~de la Macorra}
\affiliation{Instituto de F\'{\i}sica, Universidad Nacional Aut\'{o}noma de M\'{e}xico,  Circuito de la Investigaci\'{o}n Cient\'{\i}fica, Ciudad Universitaria, Cd. de M\'{e}xico  C.~P.~04510,  M\'{e}xico}

\author{P.~Doel}
\affiliation{Department of Physics \& Astronomy, University College London, Gower Street, London, WC1E 6BT, UK}

\author{S.~Ferraro}
\affiliation{Lawrence Berkeley National Laboratory, 1 Cyclotron Road, Berkeley, CA 94720, USA}
\affiliation{University of California, Berkeley, 110 Sproul Hall \#5800 Berkeley, CA 94720, USA}

\author{A.~Font-Ribera}
\affiliation{Instituci\'{o} Catalana de Recerca i Estudis Avan\c{c}ats, Passeig de Llu\'{\i}s Companys, 23, 08010 Barcelona, Spain}
\affiliation{Institut de F\'{i}sica d’Altes Energies (IFAE), The Barcelona Institute of Science and Technology, Edifici Cn, Campus UAB, 08193, Bellaterra (Barcelona), Spain}

\author{J.~E.~Forero-Romero}
\affiliation{Departamento de F\'isica, Universidad de los Andes, Cra. 1 No. 18A-10, Edificio Ip, CP 111711, Bogot\'a, Colombia}
\affiliation{Observatorio Astron\'omico, Universidad de los Andes, Cra. 1 No. 18A-10, Edificio H, CP 111711 Bogot\'a, Colombia}

\author{E.~Gaztañaga}
\affiliation{Institut d'Estudis Espacials de Catalunya (IEEC), c/ Esteve Terradas 1, Edifici RDIT, Campus PMT-UPC, 08860 Castelldefels, Spain}
\affiliation{Institute of Cosmology and Gravitation, University of Portsmouth, Dennis Sciama Building, Portsmouth, PO1 3FX, UK}
\affiliation{Institute of Space Sciences, ICE-CSIC, Campus UAB, Carrer de Can Magrans s/n, 08913 Bellaterra, Barcelona, Spain}

\author{Satya~{Gontcho A Gontcho}}
\affiliation{University of Virginia, Department of Astronomy, Charlottesville, VA 22904, USA}

\author{G.~Gutierrez}
\affiliation{Fermi National Accelerator Laboratory, PO Box 500, Batavia, IL 60510, USA}

\author{J.~Guy}
\affiliation{Lawrence Berkeley National Laboratory, 1 Cyclotron Road, Berkeley, CA 94720, USA}

\author{D.~Huterer}
\affiliation{Department of Physics, University of Michigan, 450 Church Street, Ann Arbor, MI 48109, USA}
\affiliation{University of Michigan, 500 S. State Street, Ann Arbor, MI 48109, USA}

\author{M.~Ishak}
\affiliation{Department of Physics, The University of Texas at Dallas, 800 W. Campbell Rd., Richardson, TX 75080, USA}

\author{R.~Joyce}
\affiliation{NSF NOIRLab, 950 N. Cherry Ave., Tucson, AZ 85719, USA}

\author{A.~Kremin}
\affiliation{Lawrence Berkeley National Laboratory, 1 Cyclotron Road, Berkeley, CA 94720, USA}

\author{O.~Lahav}
\affiliation{Department of Physics \& Astronomy, University College London, Gower Street, London, WC1E 6BT, UK}

\author{M.~Landriau}
\affiliation{Lawrence Berkeley National Laboratory, 1 Cyclotron Road, Berkeley, CA 94720, USA}

\author{L.~Le~Guillou}
\affiliation{Sorbonne Universit\'{e}, CNRS/IN2P3, Laboratoire de Physique Nucl\'{e}aire et de Hautes Energies (LPNHE), FR-75005 Paris, France}

\author{M.~E.~Levi}
\affiliation{Lawrence Berkeley National Laboratory, 1 Cyclotron Road, Berkeley, CA 94720, USA}

\author{M.~Manera}
\affiliation{Departament de F\'{i}sica, Serra H\'{u}nter, Universitat Aut\`{o}noma de Barcelona, 08193 Bellaterra (Barcelona), Spain}
\affiliation{Institut de F\'{i}sica d’Altes Energies (IFAE), The Barcelona Institute of Science and Technology, Edifici Cn, Campus UAB, 08193, Bellaterra (Barcelona), Spain}

\author{A.~Meisner}
\affiliation{NSF NOIRLab, 950 N. Cherry Ave., Tucson, AZ 85719, USA}

\author{R.~Miquel}
\affiliation{Instituci\'{o} Catalana de Recerca i Estudis Avan\c{c}ats, Passeig de Llu\'{\i}s Companys, 23, 08010 Barcelona, Spain}
\affiliation{Institut de F\'{i}sica d’Altes Energies (IFAE), The Barcelona Institute of Science and Technology, Edifici Cn, Campus UAB, 08193, Bellaterra (Barcelona), Spain}

\author{A.~D.~Myers}
\affiliation{Department of Physics \& Astronomy, University  of Wyoming, 1000 E. University, Dept.~3905, Laramie, WY 82071, USA}

\author{S.~Nadathur}
\affiliation{Institute of Cosmology and Gravitation, University of Portsmouth, Dennis Sciama Building, Portsmouth, PO1 3FX, UK}

\author{F.~Prada}
\affiliation{Instituto de Astrof\'{i}sica de Andaluc\'{i}a (CSIC), Glorieta de la Astronom\'{i}a, s/n, E-18008 Granada, Spain}

\author{I.~P\'erez-R\`afols}
\affiliation{Departament de F\'isica, EEBE, Universitat Polit\`ecnica de Catalunya, c/Eduard Maristany 10, 08930 Barcelona, Spain}

\author{G.~Rossi}
\affiliation{Department of Physics and Astronomy, Sejong University, 209 Neungdong-ro, Gwangjin-gu, Seoul 05006, Republic of Korea}

\author{E.~Sanchez}
\affiliation{CIEMAT, Avenida Complutense 40, E-28040 Madrid, Spain}

\author{D.~Schlegel}
\affiliation{Lawrence Berkeley National Laboratory, 1 Cyclotron Road, Berkeley, CA 94720, USA}

\author{J.~Silber}
\affiliation{Lawrence Berkeley National Laboratory, 1 Cyclotron Road, Berkeley, CA 94720, USA}

\author{D.~Sprayberry}
\affiliation{NSF NOIRLab, 950 N. Cherry Ave., Tucson, AZ 85719, USA}

\author{G.~Tarl\'{e}}
\affiliation{University of Michigan, 500 S. State Street, Ann Arbor, MI 48109, USA}

\author{B.~A.~Weaver}
\affiliation{NSF NOIRLab, 950 N. Cherry Ave., Tucson, AZ 85719, USA}

\collaboration{DESI Collaboration}

\begin{abstract}
We study the origin of extragalactic gamma-ray emission observed by Fermi-LAT, using the cross-correlation of the gamma-ray sky with maps of large-scale structure provided by the DESI and unWISE surveys. Tomographic cross-correlation reveals the bias-weighted redshift distributions of gamma-ray sources. We first illustrate this method by cross-correlating detected gamma-ray point sources with large-scale structure. We find a significant cross-correlation and infer a point source redshift distribution broadly consistent with the distribution of identified optical counterparts previously reported in the literature, as well as a similar linear bias ($b \approx 2$)
to massive galaxies that host bright active galactic nuclei. We then study the clustering of the Fermi unresolved gamma-ray background (UGRB), both in auto-correlation and in cross-correlation with large-scale structure. We detect the cross-correlation of the UGRB and LSS at $\sim 10\sigma$ in total, with highly significant detections from both DESI and unWISE. Our measurements suggest that the redshift distribution of the UGRB is broadly consistent with the redshift distribution of detected point sources. Additionally, we find a relatively weak amplitude for the cross-correlation with large-scale structure at $z < 2$, suggesting a significant fraction of the UGRB does not come from $z < 2$ large-scale structure. A natural candidate is contamination of from residual Galactic emission, and our best estimate of the contamination level derived from the UGRB auto-spectrum suggests that the mean bias of UGRB sources is indeed quite similar to the bias of detected Fermi point sources. However, we cannot exclude additional emission from gamma-ray sources at high redshift, $z > 2$, and we suggest that cross-correlation with tracers at $z > 2$, including CMB lensing, would be the ideal way to determine the fraction of $z > 2$ emission.
\end{abstract}

\maketitle

\section{Introduction}
\label{sec:intro}

The observed gamma-ray sky \cite{Ballet2023} contains both identified sources and unidentified sources.  The former class includes emission from within our own Milky Way Galaxy, as well as extragalactic sources detected individually using Fermi-LAT data \cite{2009ApJ...697.1071A, Ajello2022, atwood2013pass8realizationfermilat, 2018arXiv181011394B}, such as active galactic nuclei or star-forming galaxies.  When the Galactic plane and resolved astrophysical sources 
are removed, the remaining gamma-ray sky is called the unresolved gamma-ray background (UGRB), and the nature of the sources of the UGRB has been a topic of considerable study \cite{Ando2009,Harding2012,Tamborra2014,DiMauro2014a,DiMauro2014b,DiMauro2015,Linden2017,Korsmeier2022,Blanco2023}.  A portion of the UGRB arises from fainter examples of the known populations of detected sources, but new kinds of gamma-ray sources could also be lurking below the detection threshold, potentially even involving exotic physics such as dark matter interactions.  For this reason, it is important to characterize the nature of the UGRB, but because we cannot study these undetected sources individually, we must instead rely on more statistical probes of their nature, e.g.\ \cite{Daylan2017,Amerio2024}.
In the past decade, significant effort has been devoted to studying the UGRB through measurements of the differential source-count distribution function, $dN/dS$, which describes the number of gamma-ray sources as a function of flux $S$ \cite{Ackermann_2016,Zechlin_2016, Amerio_2023}. This quantity can be directly extracted from the Fermi-LAT sky maps using statistical methods and is related to the underlying gamma-ray luminosity function of the source populations (see Eq. 3 of \cite{Fornasa_2015}).

To probe the faint, unresolved regime (below the nominal Fermi-LAT detection threshold), \cite{Zechlin_2016} applied the 1-point PDF pixel-count statistics to Fermi-LAT data, enabling a statistical characterization of the unresolved gamma-ray source population. Applying this method to the 1–10 GeV energy range, they measured the $dN/dS$ distribution down to fluxes of  $\sim 2 \times 10^{-11} \text{ cm}^{-2} \text{ s}^{-1}$, finding a faint-end power-law index of 1.97$\pm$0.03.

More recently, \cite{Amerio_2023} applied deep-learning techniques to this problem, training convolutional neural networks on synthetic sky maps. When applied to Fermi-LAT data, these networks enabled the reconstruction of the source-count distribution down to flux levels a factor of $\sim50$ below the nominal detection threshold, showing that the unresolved regime is consistent with a scaling $dN/dS \sim S^{-2}$ down to $\sim 5 \times 10^{-12} \text{ cm}^{-2} \text{ s}^{-1}$, within the uncertainty bands.

Another statistical probe of the UGRB is its clustering on the sky, including both its autocorrelation \cite{Ackermann2012,Ackermann2018, fornasa2016angular} and its cross-correlation with other tracers of large-scale structure \cite{Xia2011,Xia2015,Branchini2017,Ammazzalorso2020,Paopiamsap2024,Thakore2025,Zhang2026,Thakore2026, shirasaki2014cross, shirasaki2016cosmological, Pinetti:2025hgd, troster2017cross, regis2015particle, Pinetti:2019ztr, cuoco2015dark, shirasaki2015cross, cuoco2017tomographic, Ammazzalorso:2018evf, shirasaki2018correlation, hashimoto2019measurement, colavincenzo2020searching, tan2020bounds, fornengo2015evidence, Arcari_2022, Fornengo_2014}.  Measurements of large-scale clustering can reveal important aspects about gamma-ray sources, even if those sources are not individually detected \cite{Zhou2024}.  A recent example of this is the cross-correlation of the UGRB with cosmic shear maps measured by the Dark Energy Survey (DES) \cite{Thakore2025}.  The cross-correlation with DES lensing was detected at high significance, and intriguingly, was found by the authors to favour a surprisingly large value for the linear bias of the gamma-ray sources.  This work reports the quantity $A_{\rm 2halo}\approx 7$, which if interpreted as the linear bias would suggest that typical sources of the UGRB reside in the most massive dark matter halos, e.g.\ galaxy clusters with virial masses $M>10^{15} \,\Msun$.  

However, one difficulty in interpreting the cross-correlation of the UGRB with weak lensing maps is the limited amount of tomographic information provided by these datasets.  To see why this poses a challenge, note that the amplitude of the cross-correlation depends on the extent of the spatial overlap between UGRB sources and the lenses probed by DES sources.  For example, if UGRB sources are at similar redshifts as DES lenses, then a stronger cross-correlation is expected than if the UGRB sources are at systematically different redshifts (either greater or lesser) than the DES lenses, holding fixed the angular clustering of the UGRB sources. The exact same cross-correlation could be obtained using weakly clustered gamma-ray sources at the same redshifts as DES lenses, or using extremely clustered gamma-ray sources at quite different redshifts than the DES lenses.  This degeneracy could be mitigated using tomography, i.e.\ by cross-correlating the UGRB with samples with precise 3D information.  DES imaging does provide some tomographic information, since multi-band photometry may be used to assign photometric redshifts to the source galaxies.  As shown in \cite{Thakore2025}, however, the DES source galaxy photo-z bins are relatively broad (with $\Delta z \sim \bar z$) and they overlap significantly, meaning that the corresponding lensing kernels also overlap quite significantly in redshift.  This makes it difficult to extract tomographic information from the cross-correlation with DES lensing, which hinders the interpretation of the measured cross-correlation.  Similarly, other works have cross-correlated Fermi UGRB maps with galaxy samples with photometric redshift estimates derived from broadband imaging \cite{Paopiamsap2024,Thakore2026}, finding consistent results.  

A relatively cleaner approach for tomographic cross-correlation is to use samples of sources with precise redshifts \cite{Newman2008,McQuinn2013,Menard2013,Newman2022}.  Cross-correlation with 3D galaxy maps allows us to extract 3D information from projected 2D datasets like the cosmic microwave background \cite{White2022,Farren2024,Farren2025} and the cosmic infrared background \cite{Jego2023,Chiang2025}, suggesting that the same method should be similarly fruitful for the UGRB.  An especially powerful tomographic dataset for these purposes has recently been provided by the Dark Energy Spectroscopic Instrument (DESI), which ultimately will measure redshifts for 63 million spectroscopically confirmed galaxies and quasars spanning the range $0<z\lesssim 2$ over an area of 17,000 square degrees on the sky  \cite{DESILyAF,DESI.DR2.BAO.cosmo,Andrade2025}.  

In this paper, we cross-correlate the Fermi-LAT $\gamma$-ray sky  with galaxy samples from  DESI data release 2 (DR2) \cite{DESILyAF,DESI.DR2.BAO.cosmo,Andrade2025}
and also from the unWISE catalog \cite{Schlafly2019,Krolewski2020}.
While not a 3D galaxy survey, the unWISE Blue, Green, and Red samples have less redshift overlap than weak lensing kernels, span a broader redshift range, and cover a much wider area on the sky, making them an advantageous sample to maximize the signal-to-noise of the cross-correlation.
  Both for point sources and for the UGRB, we detect significant cross-correlations with samples spanning a wide range of redshifts, and we then use these cross-correlations to constrain the bias-weighted redshift distribution of the gamma-ray flux, $b\,df/dz$.  
In \cref{sec:data}, we describe the various data sets used in this study.  In \cref{sec:analysis}, we describe our analysis methods, and in \cref{sec:results} we present our results.

\section{Data} \label{sec:data}

\subsection{Fermi}

\subsubsection{Point source catalog} \label{sec:pointsrccatalog}

We make use of the 4FGL-DR4 catalog of detected $\gamma$-ray point sources, compiled by the Fermi-LAT collaboration \cite{2022ApJS..260...53A, 2023arXiv230712546B, Ballet2023}. The catalog consists of 7195 spatially unresolved (point-like) sources, whose test statistic
exceeds a threshold TS $>25$, corresponding to a detection significance greater than 4$\sigma$.\footnote{Note that $\mathrm{TS}=25$ in the catalog should be compared to a $\chi^2$ distribution with four degrees of freedom, of which two related to the energy spectrum (spectral normalization and spectral index parameters), and two related to the sky direction \cite{Daylan_2017}.}  These sources are finely categorized into 24 separate types, which we aggregate together into Galactic (all sources with an identified or associated Galactic counterpart, primarily pulsars and supernova remnants), extragalactic sources (both identified and associated), and unassociated sources (consisting of both the unknown and unassociated classes). The extragalactic sources primarily consist of 
 BL Lac (blazars), FSRQ (flat-spectrum radio quasars), and BCU (blazar candidates of uncertain origin), based on association with point source detections at other wavelengths; see \cite{4FGL} for more detail.  We will study both all the extragalactic sources aggregated together, and the three most common subtypes.
For cross-correlation with the point sources, we use only the subset of sources labeled as extragalactic, neglecting galactic and unassociated sources.  

\subsubsection{Unresolved $\gamma$-ray background} \label{sec:ugrbmap}

In our analysis, we construct our own maps of the unresolved $\gamma$-ray background.  In principle, we could instead use the UGRB maps made publicly available by the previous analysis of 
\citet{Thakore2025}.  The Galactic mask used in their maps is a simple cut, removing $|b|<30 \degree$.  To test for residual Galactic contamination, we have additionally constructed independent UGRB maps with a more conservative Galactic mask, as explained in \cref{sec:ugrbcross} below. The results reported in \cref{sec:results} are primarily based on our own UGRB maps, but we have also repeated the analysis using the maps of \cite{Thakore2025} as a cross-check. 

To construct UGRB maps, we analyzed 16.5 years of Fermi-LAT data (up to week 870) using the Fermi Science Tools (\texttt{fermitools}, version 2.2.0\footnote{\url{https://github.com/fermi-lat/Fermitools-conda/wiki}}) to extract photon counts, exposure maps, and the point spread function (PSF). The data were processed in seven energy bins with boundaries at [0.5, 1, 2, 5, 10, 50, 200, 1000] GeV. For event selection, we adopted the \texttt{SOURCEVETO} class, which effectively suppresses the charged cosmic-ray (CR) background while retaining large event statistics. We used \texttt{FRONT}-converting events, having the best angular resolution. These events correspond to gamma rays detected in the upper layers of the tracker, which provides a more precise directional reconstruction. To remove atmospheric gamma-ray contamination from the Earth’s limb, we applied a maximum zenith angle cut of $Z_{\rm max} = 90^{\circ}$.

\begin{figure*}
    \includegraphics[trim={4cm 0 3cm 0},clip,width=1.05\linewidth]{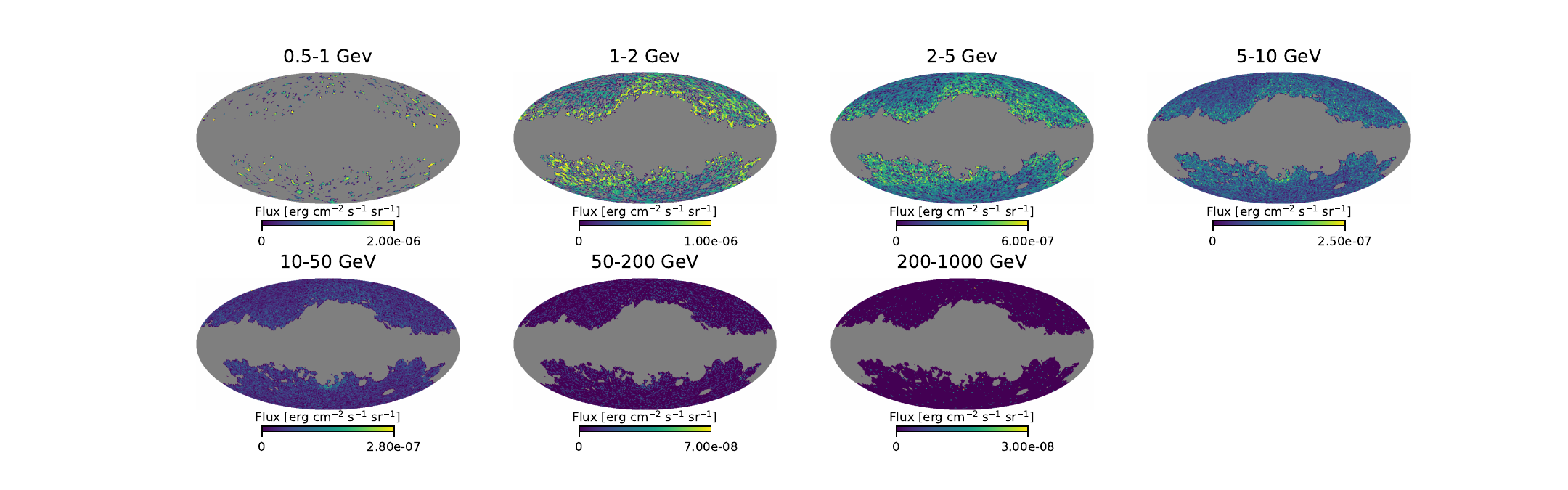}
    \caption{Fermi flux maps in each of the seven energy bins considered (see \cref{tab:fermi_sample_summary} for description). The mask that we use is shown in gray, showing the increasing mask around point sources at low energy. The map is visualized at NSIDE=64 to reduce pixel-to-pixel variance, but all analysis is performed at NSIDE=1024.}
    \label{fig:fermi_maps_raw}
\end{figure*}

\begin{figure*}
    \includegraphics[trim={4cm 0 3cm 0},clip,width=1.05\linewidth]{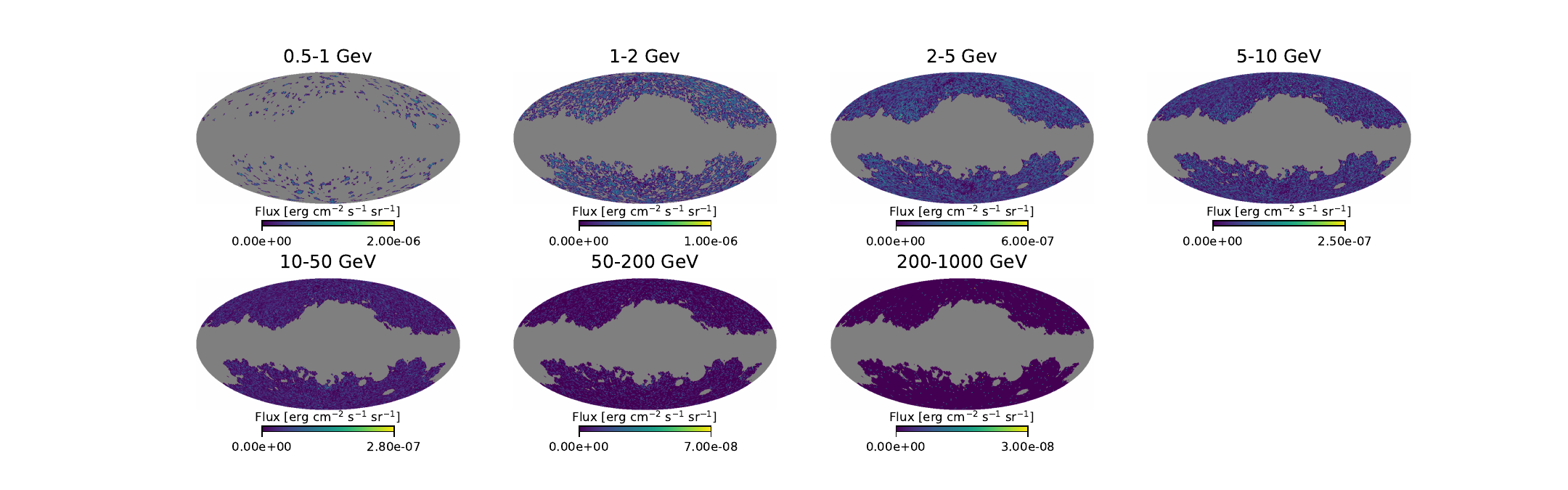}
    \caption{As in Fig.~\ref{fig:fermi_maps_raw}, but subtracting the best-fit Galactic template (a free scaling times the gll\_iem\_v07 map).}
    \label{fig:fermi_maps_galactic_subtracted}
\end{figure*}

To study the UGRB, we closely follow the analysis of the Fermi-LAT collaboration, as presented in~\cite{Ackermann2018}. The following procedure is applied independently to each energy bin:

\begin{enumerate}
    \item We begin with a photon count map generated with \texttt{fermitools} at HEALPix resolution of order 10 ($N_{\rm side} = 1024$).
    \item We mask the Galactic plane by excluding all pixels where the flux from the Galactic Interstellar Emission Model (\texttt{gll\_iem\_v7.fits}), denoted as $\mathcal{G}$, exceeds three times the isotropic background (\texttt{iso\_P8R3\_SOURCEVETO\_V3\_FRONT\_v1.txt}), denoted as $\mathcal{F}$, in the 1–2 GeV energy bin. This procedure effectively removes the bulk of the Galactic emission.
    \item We extract the positions and fluxes of \emph{all} gamma-ray sources listed in the 4FGL-DR4 Fermi catalog.
    
    \item For each energy bin [$E_{\min}$, $E_{\max}$], the point sources are masked with a circular region of radius $r_{\rm src}$, which depends on the source flux $\phi_{\rm src}$ and on the 68\% containment angle of the instrument PSF evaluated at $E_{\min}$, denoted as ${\rm PSF}(E_{\min})$. The radius is determined from the interpolation relation  
\begin{equation}
    \frac{r_{\mathrm{src}}(\phi_{\mathrm{src}}, E_{\min}) - 2 \, \mathrm{PSF}(E_{\min})}{5 \, \mathrm{PSF}(E_{\min}) - 2 \, \mathrm{PSF}(E_{\min})} 
    = \frac{\log(\phi_{\mathrm{src}}) - \log(\phi_{\mathrm{min}})}{\log(\phi_{\mathrm{max}}) - \log(\phi_{\mathrm{min}})} \, .
\end{equation}

\item Known extended sources are masked with a circular region of radius $5^\circ$. In addition, the Large Magellanic Cloud (LMC) and the lobes of Centaurus~A are masked with a radius of $10^\circ$.
The ``raw'' UGRB maps (before subtracting residual Galactic foregrounds) are in Fig.~\ref{fig:fermi_maps_raw}). The impact of the PSF-dependent masking around known sources is very apparent at low energies, where the large PSF means that most of the sky is masked, leaving an unmasked sky fraction of only $\sim$2\% in the first energy bin.

\item To subtract the residual Galactic foreground, we model the photon counts, $\mathcal{C}$, in the unmasked pixels as $\mathcal{C} = N \cdot \mathcal{G} + \mathcal{F}'$. Here, $\mathcal{G}$ is the Galactic foreground template, and $\mathcal{F}'$ represents an additional isotropic background component. We perform a joint fit for the normalization factor, $N$, and the isotropic term, $\mathcal{F}'$, in each energy bin. The resulting best-fit Galactic contribution, $N \cdot \mathcal{G}$, is then subtracted from the photon counts in the unmasked pixels.

\item The resulting masked Fermi photon count maps for the seven energy bins are shown in Fig.~\ref{fig:fermi_maps_galactic_subtracted}.
\end{enumerate}

A summary of the analysis settings is given in Table~\ref{tab:settings}.
We additionally show the PSF, the unmasked area, and the normalization of the Galactic foreground in Table~\ref{tab:fermi_sample_summary}.

\begin{table}[t]
\centering
\begin{tabular}{ | l| l|  }
\hline
Healpix order & 10, 11\\ 
Weeks & $9-870$\\ 
Emin & 500 MeV\\
Emax & 1 TeV\\
\# ${\rm E_{bins}}$ & 7 \\
Instrument Response Functions (IRFs) & \texttt{P8R3\_SOURCEVETO\_V3} \\
EVCLASS & 2048 (\texttt{SOURCEVETO})\\ 
EVTYPE & 1 (\texttt{FRONT})\\
ZMAX & $90^\circ$\\
Galactic foreground model & \texttt{gll\_iem\_v7.fits} \\
\hline
\end{tabular}
\caption{\texttt{Fermi Science Tools} settings used for the 16.5-year data set analysis.}
\label{tab:settings}
\end{table}

\begin{table*}[!ht]
\setlength{\tabcolsep}{9pt}
\centering
\begin{tabular}{l | c | c | c }
\hline
Energy bin & PSF (FWHM) & $f_{\textrm{sky}}$ & Galactic template normalization \\
\hline
0.5-1 GeV & 0.91$^\circ$ & $0.02$ & $0.94 \pm 0.013$ \\
1-2 GeV & 0.48$^\circ$ & 0.20 &$0.94 \pm 0.006$ \\
2-5 GeV & 0.27$^\circ$ & 0.37 & $0.93 \pm 0.006$ \\
5-10 GeV & 0.15$^\circ$ & 0.44 & $0.95 \pm 0.014$ \\
10-50 GeV & 0.10$^\circ$ & 0.46 & $1.00 \pm 0.022$ \\
50-200 GeV & 0.08$^\circ$ & 0.46 & $1.10 \pm 0.07$ \\
200-1000 GeV & 0.09$^\circ$ & 0.46 & $0.74 \pm 0.18$ \\
\end{tabular}
\caption{Properties of Fermi energy bins: PSF, unmasked sky fraction, and normalization of the Galactic template subtracted as shown in Fig.~\ref{fig:fermi_maps_galactic_subtracted}.
}
\label{tab:fermi_sample_summary}
\end{table*}

\subsection{DESI}

\begin{figure*}
    \includegraphics[trim={4cm 0 3cm 0},clip,width=1.05\linewidth]{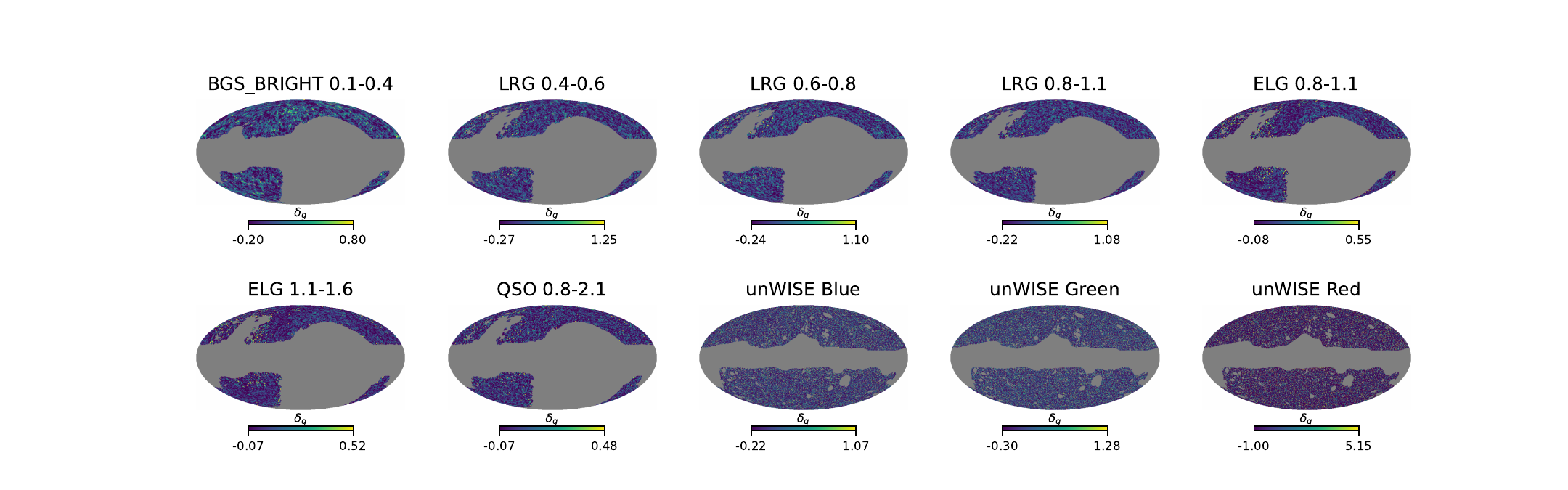}
    \caption{Galaxy overdensity maps for all 7 DESI and 3 unWISE samples used, see \cref{tab:galaxy_sample_summary} for descriptions of these samples.}
    \label{fig:galaxy_maps}
\end{figure*}

\begin{figure}
    \includegraphics[width=1.0\linewidth]{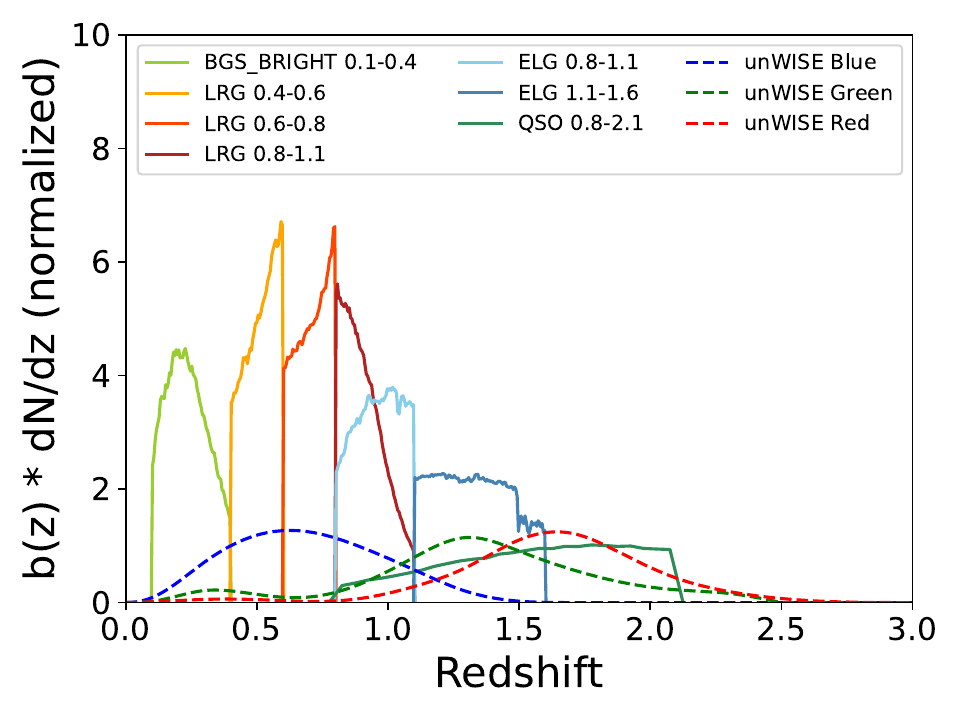}
    \caption{Bias-weighted redshift distributions for all galaxy samples used. All distributions are normalized to integrate to one.}
    \label{fig:galaxy_dndz}
\end{figure}

The Dark Energy Spectroscopic Instrument (DESI) uses a highly-multiplexed, robotically controlled focal plane to measure nearly 5000 spectra at once over a 3 deg$^2$ field from the 4-m Mayall telescope at Kitt Peak National Observatory \citep{DESI2016b.Instr,DESI2022.KP1.Instr,Corrector.Miller.2023,FiberSystem.Poppett.2024}.
DESI's eight-year survey \citep{SurveyOps.Schlafly.2023}
covering 17,000 deg$^2$ will obtain spectra of 63 million galaxies and quasars \cite{Spectro.Pipeline.Guy.2023}. The first data release (DR1) is already public \cite{DESI2024.I.DR1}, and early results are already among the most precise large-scale structure measurements available \citep{DESI2024.VII.KP7B,DESI.DR2.DR2}.

We use four different spectroscopic samples from the DESI DR2 LSS catalogs \cite{DESI.DR2.BAO.cosmo,Andrade2025,KP3s15-Ross}.  The BGS sample contains over 1 million galaxies in the redshift range $0.1<z<0.4$, the LRG sample contains over 4 million galaxies between $0.4<z<1.1$, the ELG sample contains over 6 million galaxies between $0.8<z<1.6$, and the QSO sample contains over 1 million quasars from $0.8<z<2.1$.  
We use the clustering redshift bins, BGS\_BRIGHT at $0.1 < z < 0.4$ (without the absolute magnitude cut used in the DR2 BAO sample), three LRG bins at $0.4 < z < 0.6$, $0.6 < z < 0.8$, and $0.8 < 1.1$, two ELG bins at $0.8 < z < 1.1$ and $1.1 < z < 1.6$, and one QSO bin at $0.8 < z < 2.1$. We summarize the properties of the galaxy samples in Table~\ref{tab:galaxy_sample_summary}.
We conservatively only model large scales (to $k_{\textrm{max}}$ = 0.2 $h$ Mpc$^{-1}$), and translate to a maximum angular wavenumber $\ell_{\textrm{max}}$ using the effective redshift of each sample. The magnification bias is calculated following \cite{Zhou23,WenzlChen23}. The bias evolution $b(z)$ is taken from the correlation function fits of \cite{ChaussidonY1fnl} for LRG and QSO, and we perform our own fits to BGS and ELG clustering measurements to derive the bias evolution for these tracers.

The survey selection functions, mask definitions, completeness weights, and treatment of imaging systematics are described in \cite{DESI.DR2.BAO.cosmo,KP3s15-Ross}.  DR2 covers a sky area exceeding 10,000 deg$^2$, allowing clustering to be measured on large scales in the linear regime of structure formation. In Fig.~\ref{fig:galaxy_maps}, we show sky maps of the DESI galaxies in galactic coordinates.

\begin{table*}[!ht]
\setlength{\tabcolsep}{9pt}
\centering
\begin{tabular}{l | c | c | c | c}
\hline
Galaxy sample & $z_\textrm{eff}$ & $\ell_{\textrm{max}}$ & $s$ & $b(z)$ \\
\hline
BGS\_BRIGHT 0.1-0.4 & 0.295 & 120 & 0.472 & $(1+z)^2 - 0.30$ \\
LRG 0.4-0.6 & 0.510  & 370 & 1.016 & $0.209 (1+z)^2 + 0.415$ \\
LRG 0.6-0.8 & 0.706 & 470 & 0.948  & $0.209 (1+z)^2 + 0.415$ \\
LRG 0.8-1.1 & 0.922 & 520 & 0.877  & $0.209 (1+z)^2 + 0.415$ \\
ELG 0.8-1.1 & 0.955 & 570 & 0.496  & $0.48 z + 0.822$ \\
ELG 1.1-1.6 & 1.321 & 620 & 0.609  & $0.48 z + 0.822$ \\
QSO 0.8-2.1 & 1.484 & 670 & 0.107  & $0.237 (1+z)^2 + 0.771$ \\
unWISE-Blue & 0.58 & 370 & 0.455  & -- \\
unWISE-Green & 1.26 & 570 & 0.648  & -- \\
unWISE-Red & 1.61 & 670 & 0.842  & -- \\
\hline
\end{tabular}
\caption{Effective redshift, maximum angular scale used $\ell_{\textrm{max}}$, magnification bias slope $s \equiv d\log_{10}N/dm$, and functional form of the bias evolution $b(z)$. Since we use clustering redshifts for the unWISE redshift distribution, the bias evolution is automatically included and hence we do not show a separate $b(z)$.
}
\label{tab:galaxy_sample_summary}
\end{table*}

\subsection{unWISE}

We make use of the blue, green, and red galaxy samples from the unWISE catalogue \cite{Schlafly2019,Krolewski2020}.  
These samples are selected from the W1 and W2 colors of unWISE imaging, and stars are removed based on matched Gaia \cite{Gaia_DR2} astrometry, leading to $<2\%$ stellar contamination \cite{Krolewski2020} that negligibly impacts angular clustering on the scales that we use. Imaging systematics weights remove non-cosmological fluctuations from residual correlations of stars and WISE depth with unWISE galaxies \cite{KrolewskiFerraro22,Farren2024}.
The bias-weighted redshift distributions of these samples have previously been determined by cross-correlation with SDSS spectroscopic galaxies \cite{Krolewski2020,Farren2024}. 
As in \cite{Krolewski2020,Farren2024}, we use the bias-weighted redshift distribution from clustering redshifts
when modelling Fermi cross-correlations with galaxies, and the redshift distribution from cross-matching to COSMOS photometric redshifts \cite{Laigle2015} when modelling cross-correlation with galaxy magnification.
The effective redshift and magnification bias are shown in Table~\ref{tab:galaxy_sample_summary}. We use the bias-weighted redshift distribution (from clustering redshifts) to calculate $z_{\textrm{eff}}$ for unWISE.

After masking the Milky Way, bright stars, nearby galaxies, planetary nebluae, and correcting for sub-pixel losses around fainter stars and from WISE diffraction spikes, unWISE galaxies are available on $\sim$24,000 deg$^2$ of sky, as shown in Fig.~\ref{fig:galaxy_maps}.

\newpage

\section{Analysis} \label{sec:analysis}

\subsection{Measuring angular power spectra}

\begin{figure}
    \includegraphics[width=1.0\linewidth]{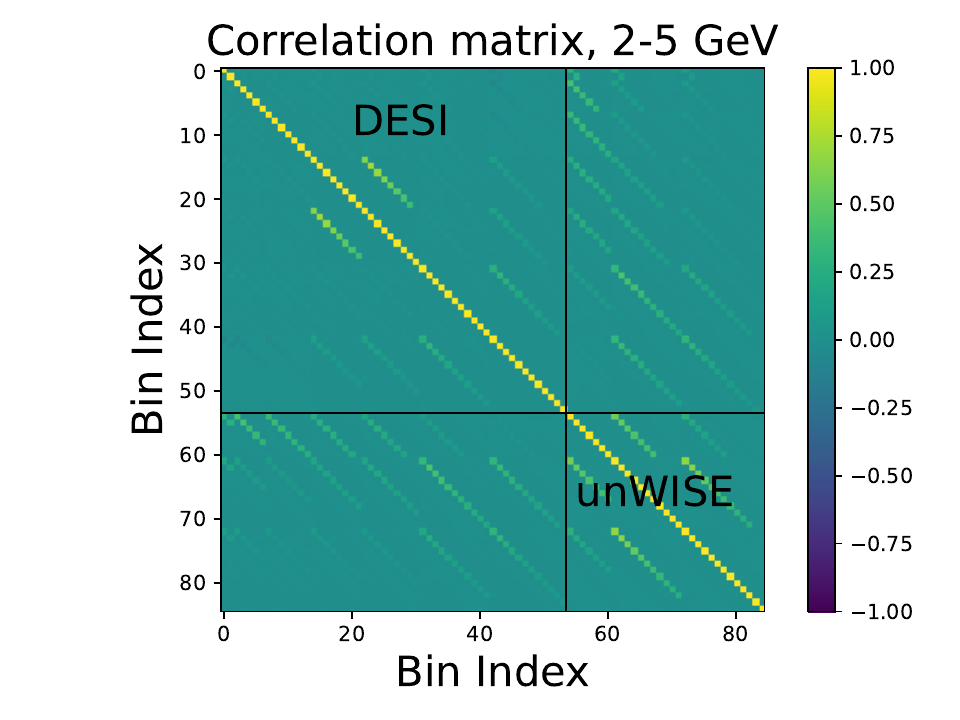}
    \caption{Correlation matrix for the 2-5 GeV Fermi background cross-correlated with DESI and unWISE. Galaxy samples are ordered following Table~\ref{tab:galaxy_sample_summary}, and the DESI and unWISE blocks are labeled. The off-diagonals come from overlapping DESI redshift bins (ELG and LRG at $0.8 < z < 1.1$, and QSO with LRG at $0.8 < z < 1.1$ and ELG at $0.8 < z < 1.6$, and between unWISE and DESI.}
    \label{fig:corr_mat}
\end{figure}

We use the MASTER algorithm \cite{Hivon2002} as implemented in the public code \texttt{NaMaster}\footnote{\url{https://github.com/LSSTDESC/NaMaster}} \cite{Alonso2019} to deconvolve the mask and measure the binned pseudo-$C_\ell$ bandpowers. We use the pixel-free estimator of \cite{Baleato2024} as implemented in \texttt{NaMaster} \cite{Wolz2025}. We calculate analytic Gaussian covariances using the implementation in \texttt{NaMaster} \cite{Garcia-Garcia2019}, including all off-diagonal covariances between different galaxy samples, as the different samples have nonzero correlation (especially when the redshift bins overlap). We use the analytic covariances for the galaxy auto-correlation, the galaxy-gamma ray cross-correlation, and the gamma-ray auto-correlation.
We use a smooth spline fit to the measured galaxy and Fermi autocorrelations as input to the analytic covariances; this fit allows for excess power in the Fermi autocorrelations at low $\ell$ above the expected signal. We show the correlation matrix for a representative bin of the unresolved background in Fig.~\ref{fig:corr_mat}.

We use broad bins of width $\Delta \ell = 50$ with $\ell_{\textrm{min}} = 20$. We measure power spectra out to the smallest scales allowed by the NSIDE=1024 Fermi maps ($\ell_{\textrm{max}} = 3072$), and conservatively model only the large scales with $k_{\textrm{max}} = 0.2$ $h$ Mpc$^{-1}$ at the effective redshift of each galaxy sample.
Because of the large Fermi beam, these conservative scale cuts do not lose much signal-to-noise over more aggressive scale cuts.

We check the validity of the mode deconvolution on simple  Gaussian mocks, generating 1000 correlated mocks with a Poisson-sampled galaxy field and a continuous field representing the Fermi maps. We find excellent agreement between the input and measured power spectra for all bins except the 0.5--1 GeV bin (with no detectable deviation between true and measured power spectrum at the level of the standard error of the mean). For the lowest-energy bin, power spectrum recovery is poor due to the very small unmasked fraction and the dispersed nature of the mask. We therefore omit this bin from all further analysis, and note that the cross-correlation is detected at $<2 \sigma$ anyway at these energies.

Since we mask the UGRB around all resolved Fermi sources, and many of the resolved sources at high latitude are extragalactic, we may introduce a bias into our analysis via
mask-map coupling \citep[e.g.][]{Surrao23}. This bias is expected to be worst for cross-correlations between the quasars (which are the most likely to be Fermi point sources) and the lower energy bins where the PSF and thus the point-source mask is largest.
We test for this bias in two ways.
First, we compare the quasar autocorrelation across the entire DESI footprint to the quasar autocorrelation when also applying the 1-2 GeV mask. We find these autocorrelations agree to 10\%, and the remaining fluctuations are consistent with noise given the different sky areas covered. This suggests that our measured power spectra will not be substantially impacted by map-mask coupling.
Second, we created a synthetic mask out of the quasar catalog. We randomly picked quasars at the density of the Fermi point sources, and masked 1.5$^{\circ}$ around them; this produced a mask that visually agrees with the 1-2 GeV bin mask, and matches the $f_{\textrm{sky}}$ to within 10\%. We then compared the original quasar autocorrelation to the quasar autocorrelation when also applying this synthetic mask. This acts as a maximal test of map-mask correlation (since not all of the Fermi point sources are $0.8 < z < 2.1$ DESI quasars, as some are optically faint or at the wrong redshifts). In this case, we find 5\% variations in the power spectra, again entirely consistent with the fluctuations expected due to the different sky area. We therefore conclude that the impact of mask-map coupling on our cross-correlation measurements is subdominant to their statistical errors.

\subsection{Modelling redshift distribution and power spectra}
\label{sec:dndzmodel}

Because the cross-correlation with the Fermi point sources and unresolved background is detected with only moderate signal-to-noise ratio, we cannot reconstruct the bias-weighted redshift distribution in a model-independent way.  Instead, we write down a parametric model for $b\, dN/dz$, and fit the model's parameters to the measured cross-spectra.  Our adopted model takes the form
\begin{equation} \label{eq:dndz}
    b\frac{dN}{dz} \propto z^2 \exp{\left(-\left(\frac z {z_\star}\right)^\beta\right)},
\end{equation}
where the free parameters are $z_\star$, $\beta$, and the overall normalization. 
This is a simple, well-motivated functional form used in weak lensing studies and in the BPZ photometric redshift method \cite{Benitez98,Wittman00}.
We impose a flat prior on $z_\star$ between 0 and 3, on $\log_{10}(\beta)$ between -2 and 2, and on the derived parameter $z_{\textrm{peak}}$ (the maximum of the redshift distribution) between 0 and 2.1. These priors are chosen to favor the redshift range over which we have galaxy samples ($0 < z < 2$) and we thus caution that we cannot constrain the possibility of a second peak or $z>2$ contribution to the redshift distribution.
The priors chosen lead to a flat prior distribution of $z_{\textrm{peak}}$, and a flat distribution of mean redshift $\bar{z}$ at $0 < \bar{z} < 1.5$, with the prior distribution dropping to higher $\bar{z}$.

We use a simple model for the angular power spectra, a linear bias times the nonlinear matter power spectrum
\cite{Mead20}. On the scales that we use ($k < 0.2$ $h$ Mpc$^{-1}$), the inaccuracies in the modelling are much smaller than the uncertainty in our data.  This includes not only neglect of nonlinear bias, but also neglect of shot noise in the cross-correlations. This is justified because the cross-shot noise only potentially becomes significant compared to the 2-halo term on scales close to $\ell_{\textrm{max}}$, where the signal-to-noise in the measured cross-spectra is quite small (see \cref{sec:corr_plots}).

\section{Results} \label{sec:results}

\subsection{Fermi point sources} \label{sec:pointsrc}

\begin{figure}
    \includegraphics[width=0.98\linewidth]{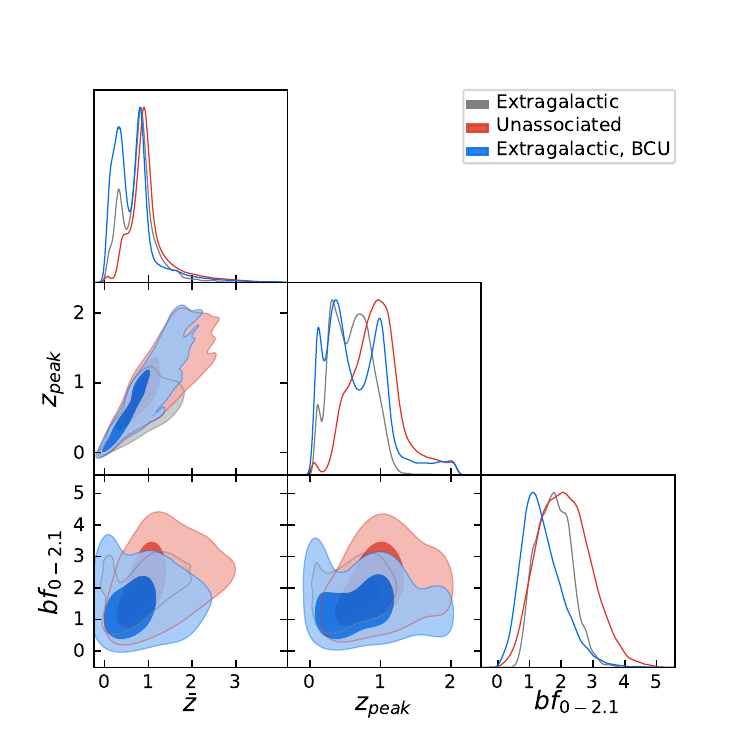}
    \caption{Constraints on redshift distribution and product of the Fermi bias times the fraction at $0 < z < 2.1$, for extragalactic Fermi point sources (gray), sources without a confirmed counterpart and redshift (red), and ``blazar candidates of uncertain origin'' (BCU), a subset of the extragalactic sources with especially uncertain $dN/dz$ due to a smaller fraction of optical counterparts compared to confirmed blazars or flat-spectrum radio quasars.
    }
    \label{fig:pointsourcecorner}
\end{figure}

\begin{figure}
    \includegraphics[width=0.98\linewidth]{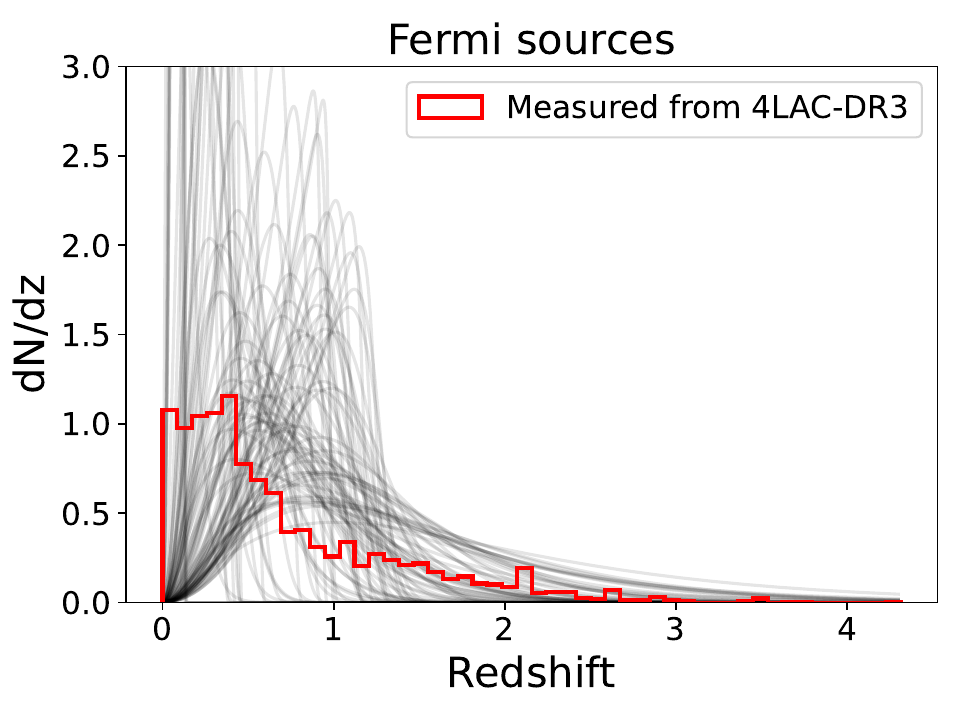}
    \caption{Samples of Fermi source $dN/dz$ (gray lines) compared to measured redshift distribution from optical counterparts (red histogram).}
    \label{fig:pointsourcedndz}
\end{figure}

\begin{table*}[!ht]
\centering
\begin{tabular}{l | c c c c c c}
\hline
Point source sample & $z_{\textrm{peak}}$ & $bf_{\textrm{0 - 2.1}}$ & Bias &  $\chi^2$ & $p$-value & Det.\ sig.\ ($N\sigma$) \\
\hline
Extragalactic & $0.6 \pm 0.3$ & $1.8^{+0.5}_{-0.6}$ & -- & 83.2 & 0.44 & 6.0 \\
Unassociated & $1.1^{+0.4}_{-0.7}$ & $2.1^{+0.8}_{-0.9}$ & -- & 65.9 & 0.90 & 3.6 \\
Extragalactic, BCU & $0.98^{+0.4}_{-1.0}$ & $1.4^{+0.5}_{-0.8}$ & -- & 68.0 & 0.87 & 3.3 \\
\hline
Extragalactic & -- & -- & $2.4 \pm 0.4$ & 86.4 & 0.35 & 5.7 \\
Extragalactic, BL Lac & -- & -- & $0.9^{+1.3}_{-0.6}$ & 124.4 & 0.0018 & 0.0 \\
Extragalactic, FSRQ & -- & -- & $2.5^{+0.8}_{-0.9}$ & 79.0 & 0.57 & 3.0\\
\hline
\end{tabular}
\caption{Marginalized constraints on the peak redshift and bias times fraction between $z=0$ and $z=2.1$ for the Fermi point sources.
The $\chi^2$ is for the best-fit parameters and we give the corresponding $p$ value for 82 degrees of freedom (85 data-points and 3 free parameters). 
In the top three rows, both the bias and the $b \, dN/dz$ parameters are varied; in the bottom three rows, $dN/dz$ is fixed using the Fermi sources with matching redshifts, and only the linear bias is varied.
Detection significance is defined as $\sqrt{\vec{d} \, \mathbf{C}^{-1} \, \vec{d} - \chi^2}$, where $\vec{d}$ is the data vector and $\mathbf{C}$ is the covariance.}
\label{tab:point_sources_results}
\end{table*}

We first present the cross-correlation of the galaxy maps described above with the 4FGL-DR4 catalog of detected point sources described in \cref{sec:pointsrccatalog}.  This not only illustrates how cross-correlation may be used to derive tomographic properties of a population of sources, but it also is (to our knowledge) the first time clustering redshifts have been derived for Fermi point sources.  
Since our adopted galaxy samples have known bias $b_{\rm gal}$ and known redshift distributions $dN_{\rm gal}/dz$, this cross-correlation may be used to reveal the bias-weighted redshift distribution $b\,dN/dz$ of the detected point sources.  A subset of these point sources have previously been cross-matched with known galactic or extragalactic sources \cite{Ajello2022}.
We will subsequently compare the redshift distribution of that subset with our reconstructed $b\,dN/dz$ to validate our redshift distribution measurement.

\Cref{fig:pointsourcecross} in \cref{sec:corr_plots} shows the cross-spectra we measure using the sources labeled as extragalactic in the 4FGL-DR4 catalog.  Combining all galaxy bins, we detect cross-correlation with large-scale structure at $\approx 6\,\sigma$ confidence.

\Cref{fig:pointsourcecorner} shows posteriors for  parameters describing the point source population, including both the shape and normalization of $b\, dN/dz$.  If we think of $b\, dN/dz$ as the product of two functions, the bias $b(z)$ and the redshift distribution $dN/dz$, and if we require that $dN/dz$ integrates to give the total number $N$ of detected point sources, then the integral of $b\, dN/dz$ over all redshifts may be interpreted as $N$ times the mean bias of the detected point sources.  Another derived quantity of interest is $b\,f_{0-2.1}$,  the integral of $b\, dN/dz$ over the redshift range $0<z<2.1$ probed by our galaxy sample, divided by the total number of sources.  This quantity can differ from the average effective $b$ when the redshift distribution of the Fermi point sources extends significantly beyond the range covered by our galaxy samples.  For the point sources, however, the posterior disfavours redshift distributions with support extending to high redshift ($z>2$), and so we find that $b \approx b\,f_{0-2.1}$.

\Cref{fig:pointsourcedndz} shows samples of $b\, dN/dz$ drawn from this posterior, compared to the redshift distribution of the subset of point sources previously cross-matched to redshifts in the 4LAC-DR3 catalog \cite{Ajello2022}.  The point sources comprise several subsets, primarily BL Lacs, FSRQ (flat-spectrum radio quasars), and BCU (``blazar candidates of unknown origin'').
The fraction of these point sources with an optical association and thus a confident redshift varies by class: 99.5\% of the FSRQ have a redshift, 64.3\% of the BL Lacs, and 17.7\% of the BCU. Since the redshift distributions of the classes are substantially different (see Fig.~2 in \cite{Ajello2022}), we weight point sources in each sample by the inverse of that sample's redshift success rate when computing the overall $dN/dz$. We thus assume that sources without a redshift have the same $dN/dz$ as sources of the same class with a redshift.

The posterior samples appear broadly similar in shape to the measured $dN/dz$ of the 4LAC-DR3 sources, suggesting that the subset with identified counterparts appears to be representative of the full population, and that there is not extremely strong redshift evolution of the bias $b(z)$.  For example, the median (bias-weighted) redshift is $\bar z \approx 0.7$, similar to the median redshift of the sources with identified counterparts.  

As mentioned above, the normalization of the fitted $b\, dN/dz$ may be interpreted as an average bias for the point source population.  We find $\bar b \sim 2$, similar to the bias of DESI luminous red galaxies.  This suggests that the Fermi point sources occur in halos of mass similar to LRG hosts, e.g.\ $M \sim 10^{13} M_\odot$. Since active galactic nuclei (AGN) are believed to comprise a significant fraction of Fermi point sources, this bias value appears reasonable and is consistent with previous auto-correlation results \cite{Allevato14}. 

In addition to the extragalactic sources, we have repeated the above analysis for point sources labeled in the catalog as unassociated; for these sources, cross-correlations provide the only viable way to constrain the redshift distribution.  The cross-correlation is detected at somewhat lower significance, $\sim 3.6\sigma$, and so the bias-weighted redshift distribution is not constrained as tightly as it is for the extragalactic sources. Overall, we do not detect any significant differences between the redshift distributions of the extragalactic and unassociated sources, perhaps due to the uncertainties on the model parameters. The similar value of $b f_{\textrm{0-2.1}}$ between the unassociated and extragalactic sources suggests that the unassociated sources have little Galactic contamination and are dominated by extragalactic sources---this is reasonable considering that $\sim$3\% of associated sources are Galactic at $|b| > 30^{\circ}$. 

We similarly repeated this analysis for the BL Lacs, FSRQ, and BCU samples.  We detect cross-correlations with the BCU at 3.3$\sigma$, the FSRQ at 3.0$\sigma$, and do not detect cross-correlations with the BL Lacs. Because the BCU have a low redshift success rate, we show constraints on their redshift distribution in Fig.~\ref{fig:pointsourcecorner} along with all the extragalactic point sources and the unassociated point sources.


The resulting parameter constraints are summarized in \cref{tab:point_sources_results}.

\subsection{Unresolved $\gamma$-ray background} \label{sec:ugrb}

Having illustrated tomographic cross-correlations using the point source catalog, we next apply the same analysis to the unresolved gamma-ray background (UGRB) described in \cref{sec:ugrbmap}.  

\subsubsection{Auto-spectrum} \label{sec:autocorr}

Before discussing the cross-correlation of the UGRB with large-scale structure, we first consider the auto-correlation of our UGRB maps.  As discussed in \cref{sec:ugrbmap}, we have subtracted best-fit templates for diffuse Galactic emission from the UGRB maps.  If this subtraction successfully removed all Galactic emission, then the auto-spectrum of the remaining UGRB map should resemble a shot noise spectrum, simply because the number of $\gamma$-ray photons is small enough that Poisson noise should dominate over  cosmological large-scale structure for any plausible bias for the UGRB sources.   

\cref{fig:Clff1,fig:Clff2,fig:Clff3,fig:Clff4} in \cref{sec:corr_plots} show our measured auto-spectra.  
Consistent with previous analyses \cite{Ackermann2018}, we find considerable power in the auto-spectra on large angular scales, $\ell<50$, indicative of significant contamination from Galactic emission originating within the Milky Way.  
The significant contamination of the UGRB arising from Galactic emission poses a challenge for any correlation analysis.  If diffuse $\gamma$-ray emission from the Galaxy is uncorrelated with observed maps of galaxy density, this contamination would not change the shape of $b\,df/dz$ measured below, but it will affect the normalization, which is used to infer the typical host halo bias.  To infer the mean bias of the UGRB, we therefore require an estimate of what fraction of the unresolved emission in our UGRB maps arises from residual Galactic contamination.  In \cref{sec:contamination}, we use the UGRB auto-spectrum shape to estimate the contamination fraction of the UGRB, finding that $\sim 60\%$ of the $\gamma$-ray flux in our UGRB maps arises from Galactic contamination (see \cref{tab:auto_results}).  
We stress that this estimate for the contamination fraction is extremely crude, but nevertheless, having even an order-of-magnitude estimate is helpful in interpreting the cross-correlation results shown below.

\subsubsection{Cross-correlations with galaxy samples}
\label{sec:ugrbcross}

\begin{figure*}
    \includegraphics[width=0.49\linewidth]{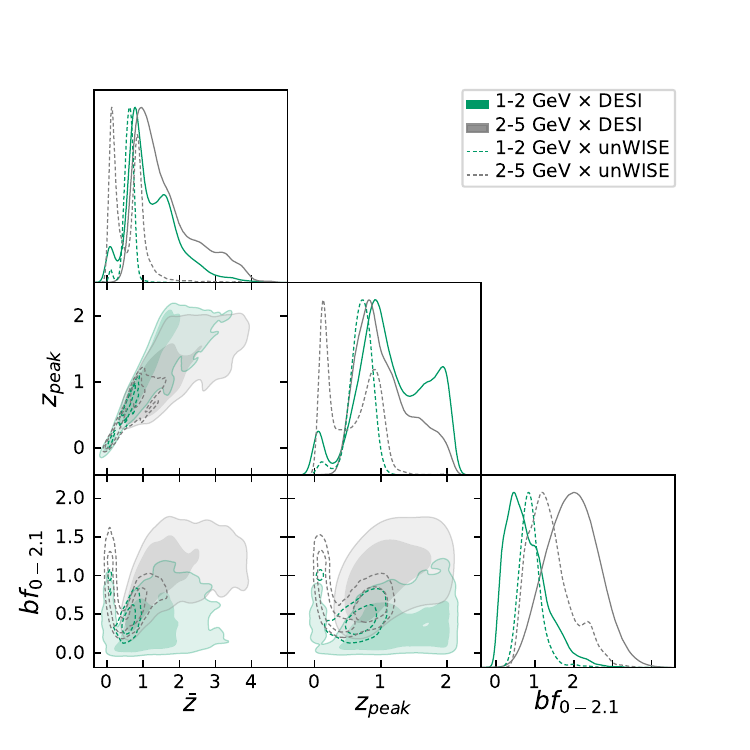}
    \includegraphics[width=0.49\linewidth]{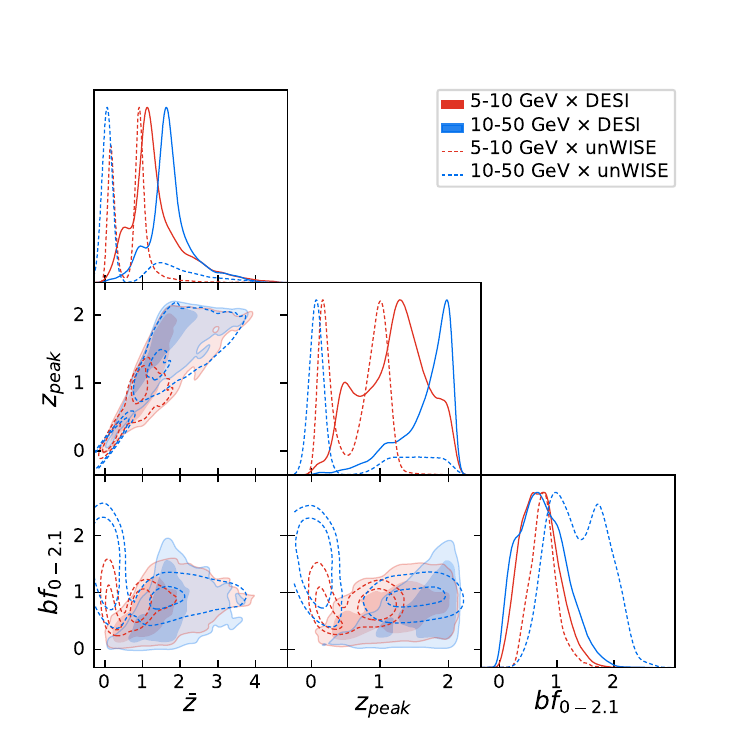}
    \caption{Comparison of marginalized parameters for cross-correlations between Fermi and DESI (solid) and Fermi cross unWISE (dashed). Left panel shows the 1--2 GeV and 2--5 GeV bins, and the right panel shows 5--10 GeV and 10--50 GeV. }
    \label{fig:compare_unwise_vs_desi}
\end{figure*}

\begin{figure}
    \includegraphics[width=0.98\linewidth]{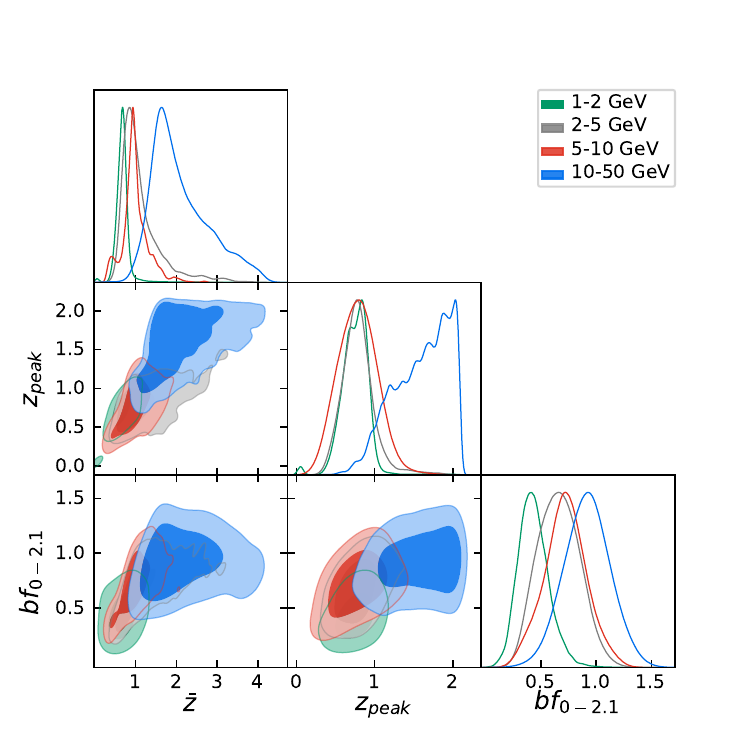}
    \caption{Comparison of constraints on the redshift distribution and the product of the Fermi UGRB bias times the fraction at $0 < z < 2.1$, for the four UGRB energy bins with the best detection significance.  Here, $\bar z$ refers to the mean (bias-weighted) redshift of the $b\,df/dz$ distribution, $z_{\rm peak}$ is the location of the maximum of $b\,df/dz$, and $b\,f_{0-2.1}=\int_0^{2.1} dz\, b\, df/dz$.}
    \label{fig:ugrb_corner}
\end{figure}

\begin{figure*}
    \includegraphics[width=0.98\linewidth]{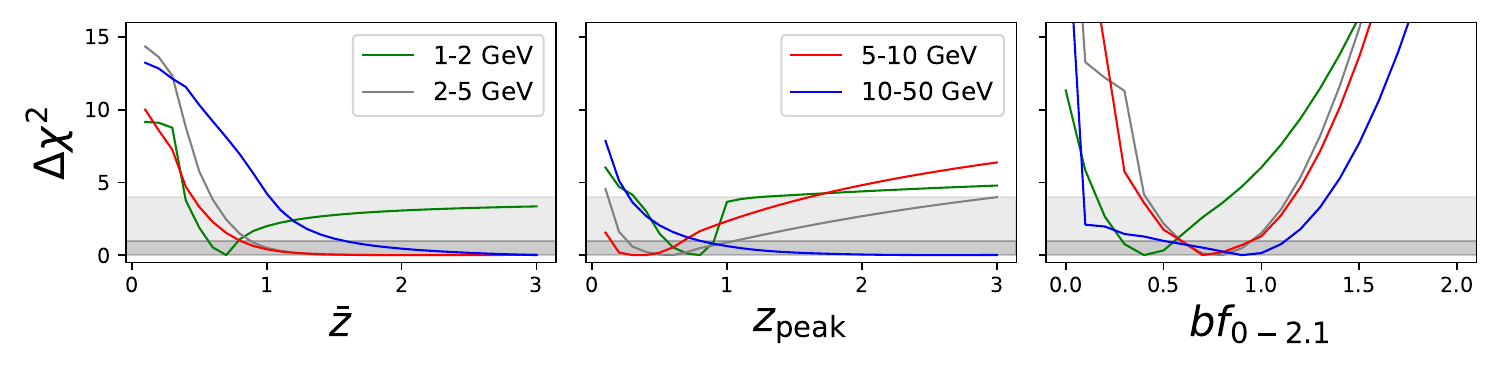}
    \caption{Profile likelihoods for the Fermi UGRB parameters for the four energy bins with the best detection significance.
    }
    \label{fig:ugrb_profile}
\end{figure*}

\begin{figure*}
    \includegraphics[width=0.98\linewidth]{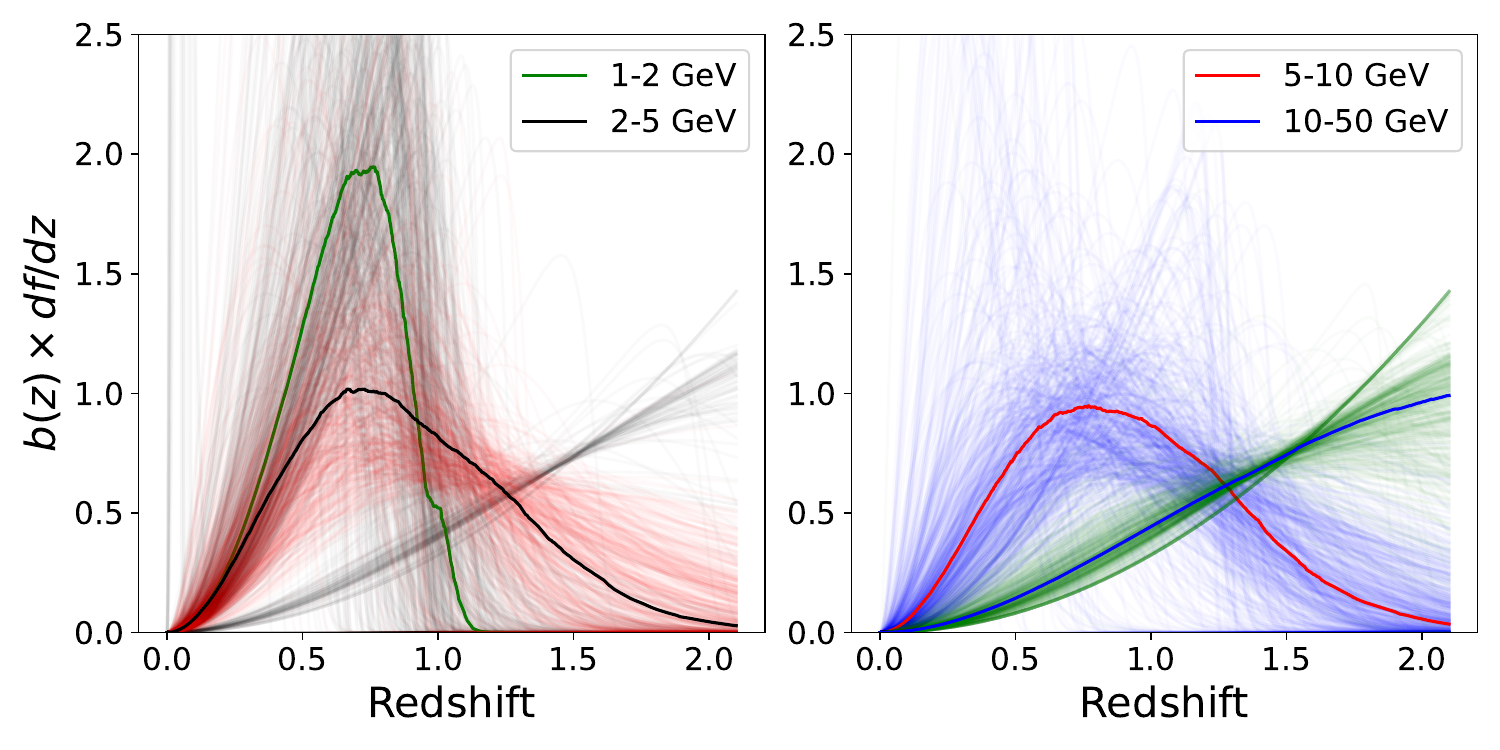}
    \caption{Samples of $b(z) \times df/dz$ for the four best-constrained Fermi energy bins, normalized to integrate to unity over $0 < z < 2.1$. The solid lines give the mean of the samples.  }
    \label{fig:summary_dndz}
\end{figure*}

We next turn to the cross-correlation analysis of the UGRB.
We cross-correlate each of the seven $\gamma$-ray maps with each of the galaxy maps, to constrain the bias-weighted redshift distribution for each energy bin.  \cref{fig:bin_1_xcorr,fig:bin_2_xcorr,fig:bin_3_xcorr,fig:bin_4_xcorr} show examples of the resulting cross-spectra.  We do not detect significant cross-correlations with large-scale structure for either the lowest or highest energy bins, due to low sky area and low photon counts, respectively (see \cref{fig:fermi_maps_galactic_subtracted}).  But for the maps with energies 1-50 GeV, we do detect significant cross-correlations.  

From these cross-correlations, we can constrain the bias-weighted redshift distribution of the unresolved flux, $b\,df/dz$.  Here, we will consider $f$ to be the fraction of flux, i.e.\ $\int dz \, df/dz = 1$,
where the Galactic template is always subtracted from the flux (i.e.\ we use maps in Fig.~\ref{fig:fermi_maps_galactic_subtracted}), but we do not correct the flux by the contamination fraction estimated in Sec.~\ref{sec:autocorr}.
As in \cref{sec:pointsrc}, the signal-to-noise of the detected cross-correlations is not sufficiently large to permit us to reconstruct $b\,df/dz$ non-parametrically.  We therefore model $b\,df/dz$ using the same parametric model given in \cref{eq:dndz}, with the same parameters $z_\star$ and $\beta$ describing the shape of $b\,df/dz$, along with the overall normalization.  We adopt the same prior on the parameters of this model for the redshift distribution that we used for the point source distribution, described in \cref{sec:dndzmodel}.  Some additional useful derived parameters characterizing the redshift distribution are the mean redshift ${\bar z} = (\int dz\, z\, b\, df/dz)/(\int dz\, b\, df/dz)$ and the peak redshift $z_{\rm peak}$ where $b\,df/dz$ is maximized.

Our measurements are consistent whether we use only unWISE data, only DESI data, or their combination, except perhaps mild tension for the 10--50 GeV bin (see Fig.~\ref{fig:compare_unwise_vs_desi}).
The overall detection significance  of $\sim$ 10$\sigma$ (obtained from the quadrature sum of the individual bin detection significances, assuming independence) is dominated by the unWISE cross-correlation (total detection 10.7$\sigma$), but we also detect a significant cross-correlation with DESI alone (6.7$\sigma$). Note that because the best-fit parameters are slightly different between the unWISE-only and DESI-only results, the unWISE-only detection significance is slightly higher than the combined detection significance; there is also a significant covariance between the unWISE and DESI detection significances due to the common sky area and redshift range. Our covariance matrices always correctly account for this covariance by including the off-diagonal covariance between different galaxy samples cross-correlated with Fermi.

\cref{tab:ugrb_results} lists the detection significances and resulting parameter constraints on the bias-weighted redshift distributions for each of these energy bins, and \cref{fig:ugrb_corner,fig:ugrb_profile} show the resulting constraints on these model parameters, as well as for derived properties describing the UGRB.  

\begin{table*}[!ht]
\centering
\begin{tabular}{l | c c c  c c c}
\hline
\multirow{2}{*}{Energy bin} & \multirow{2}{*}{$z_{\textrm{peak}}$} & \multirow{2}{*}{$bf_{\textrm{0 - 2.1}}$} & $b\, I_{\textrm{0 - 2.1}} $ &  \multirow{2}{*}{$\chi^2$} & \multirow{2}{*}{$p$-value} & Det.\ sig.\ \\
& & & {\scriptsize [erg cm$^{-2}$ s$^{-1}$ sr$^{-1}$]} & & & ($N\sigma$) \\
\hline
1--2 GeV & $0.77^{+0.14}_{-0.17}$ & $0.42^{+0.14}_{-0.13}$ & $1.65^{+0.55}_{-0.51} \times 10^{-7}$ &
103.0 & 0.06 & 3.4 \\
2--5 GeV & $0.78^{+0.17}_{-0.18}$ & $0.65\pm0.2$ & $1.15 \pm 0.4 \times 10^{-7}$ &

93.3 & 0.19 & 5.7 \\
5--10 GeV & $0.79^{+0.24}_{-0.27}$ & $0.72\pm0.2$ & $3.60 \pm 1.0 \times 10^{-8}$
& 102.3 & 0.06 & 5.5 \\
10--50 GeV & $2.49^{+1.19}_{-0.84}$ & $0.87^{+0.23}_{-0.24}$  & $2.58^{+0.68}_{-0.71} \times 10^{-8}$
& 84.4 & 0.41 & 5.0 \\
50--200 GeV & $2.35^{+1.78}_{-1.07}$ & $0.74^{+0.71}_{-0.53}$ &
$1.72^{+1.64}_{-1.23} \times 10^{-9}$
& 78.2 & 0.6 & 1.4 \\
200--1000 GeV & $2.25^{+0.97}_{-0.78}$ & $1.42^{+1.83}_{-1.13}$ & $2.43^{+3.13}_{-1.93} \times 10^{-10}$
& 73.1 & 0.75 & 0.0 \\
\hline
\end{tabular}
\caption{Marginalized constraints on the peak redshift and bias times fraction between $z=0$ and $z=2.1$ for the Fermi unresolved $\gamma$-ray background. We also show the bias-weighted mean UGRB intensity over $0 < z < 2.1$, $b I_{\textrm{0 - 2.1}} $ (in physical units).
}
\label{tab:ugrb_results}
\end{table*}

Given the large uncertainties on the model parameters, the shapes of the redshift distributions appear broadly consistent with the redshift distribution of the point sources, with similar median redshifts for the best-constrained energy bins.  One possible exception may be the 10-50 GeV energy bin, which appears to favour higher values for $z_{\rm peak}$ and $\bar z$ than the other bins.  This preference for higher redshifts is only at the $\sim 1\sigma$ level, though, so we cannot claim significant evidence for redshift evolution of the UGRB source population, e.g.\ harder sources becoming more important at higher redshift.\footnote{We note that such a trend could qualitatively be consistent with blazars dominating above 10 GeV, while active galactic nuclei or star-forming galaxies dominate below 10 GeV \cite{cuoco2017tomographic}. However, given the low statistical significance of the observed difference, we do not attempt any further interpretation.}

One notable difference between the UGRB and point sources is that the amplitude of $b\,df/dz$ for the UGRB appears to be significantly smaller than that derived for the point sources.  This can be seen from the values of $b\,f_{0-2.1}$ given in \cref{tab:ugrb_results}, which fall in the range $\sim 0.4-0.9$, significantly below the effective bias found for point sources, $b\sim 2$. 
This low amplitude of the cross-correlation is surprising, especially for the lower energy bins where the bias appears to be significantly smaller than 1.  This is difficult to reconcile with the expected bias for any plausible astrophysical sources, which all should occur within dark matter halos.  The lowest bias for even the lowest-mass halos (e.g., $M<10^8\, \Msun$) is typically $b\gtrsim 0.6$ \cite{Tinker2010}, so the inferred $b\,f_{0-2.1} \approx 0.4$ found for the 1-2 GeV bin cannot be explained by a low halo bias alone.  Instead, this suggests that a significant fraction of the UGRB arises from sources at redshifts not covered by DESI+unWISE galaxies, i.e.\ either foreground (Milky Way) or background ($z>2$) emission.  

Of these two possibilities, let us first consider Milky Way contamination.
As mentioned in \cref{sec:ugrbmap}, our UGRB maps and the maps made publicly available by \citet{Thakore2025} exhibit clear evidence for Galactic contamination, as seen in the low-$\ell$ auto-spectra of the UGRB.  This contamination provides a natural explanation for our low inferred bias, since a significant fraction of gamma-ray flux from the Galaxy would be uncorrelated with cosmological large-scale structure, reducing the cross-correlation coefficient with our galaxy maps, and hence lowering the inferred bias.  As described in \cref{sec:contamination} and \cref{tab:auto_results}, an extremely crude estimate for the fraction of $\gamma$-ray flux arising from Galactic contamination is $\sim 60\%$.  If this estimate is correct, then our measured value $b\,f_{0.2.1} \sim 0.4-0.9$ would translate into inferred bias values $b \sim 1-2$, with most of the energy bins being closer to $b\approx2$ than $b\approx 1$.  These bias values are quite similar to what we measure for the detected Fermi point sources in \cref{sec:pointsrc}, suggesting that the unresolved sources largely originate from a population similar to the resolved extragalactic sources.

The other possibility besides Milky Way contamination is high-redshift gamma-ray emission.  Our inferred $b\, f_{0.2.1}$ would be diminished if a significant fraction of the UGRB map arises from sources beyond the redshift range covered by our galaxy samples, at $z>2$.  Our measurements cannot exclude this possibility.  The posterior contours shown in \cref{fig:ugrb_corner} appear to disfavour redshift distributions peaking at $z>2$, but this appears to be a consequence of our assumed priors on the parameters describing $b\,df/dz$.  If we instead consider the profile likelihoods shown in \cref{fig:ugrb_profile}, it is apparent that significant emission from $z>2$ is not disfavoured by our measurements.  A conclusive determination of the fraction of the UGRB arising from high redshift will require further cross-correlations with samples at even higher redshift than probed by the QSO sample.  One example of a tracer that is sensitive to large-scale structure at $z>2$ is CMB lensing, which has an extremely broad kernel with significant support at high redshift.  Given that convergence maps from several CMB surveys are publicly available, this appears to be a promising route towards understanding the anomalously low cross-correlation that we observe for the Fermi UGRB.

\section{Discussion}

We have used tomographic cross-correlation to study the nature of the sources of the extragalactic gamma-ray sky.  We correlated maps of Fermi point sources and also maps of the unresolved gamma-ray background, with 3D maps of cosmological large-scale structure provided by the DESI and unWISE surveys.  Our results suggest that the UGRB and the point source population have broadly similar redshift distributions over the range $0<z\lesssim 2$ probed by our galaxy samples.  Intriguingly, we find significantly different amplitudes for the cross-correlations.  For Fermi point sources, the best-fit linear bias is of order $b\sim 2$, suggesting that these point sources occur in massive dark matter halos ($M\sim 10^{13}\Msun$), similar in mass to the halos hosting luminous red galaxies.  For the UGRB, we find significantly weaker correlations, with $b\ f_{0-2.1}\sim 0.4-0.9$, depending on gamma-ray energy.  If $f_{0-2.1}\approx 1$, then the inferred bias of the UGRB would be much smaller than the bias of detected Fermi point sources.

The latter interpretation appears to be in tension with the conclusions of recent work cross-correlating the Fermi UGRB with cosmic shear \cite{Thakore2025} and with DES galaxies \cite{Thakore2026}.  Although that work does not report a linear bias, they do report a parameter $A_{\rm 2halo}$, which is proportional to the linear bias with an unknown normalization. They find $A_{\rm 2halo}\approx 6.6^{+0.1}_{-2.2}$, which they interpret as evidence for large halo masses, similar to massive galaxy clusters which are highly biased at all redshifts \cite{Thakore2025}. Using the same UGRB map analyzed in \cite{Thakore2025}, we obtain a much lower effective bias, e.g.\ $b\lesssim 1$ (for $f_{0-2.1}\approx 1$), typical of halos less massive than the Milky Way. However, it remains unclear whether this reflects a discrepancy in the underlying shear and galaxy cross-correlation measurements themselves, or instead a difference in the interpretation of those measurements.
In the same way that galaxy cross-correlations measure only the product of $b\,df/dz$, shear cross-correlations also are sensitive to $b\,df/dz$ over the redshift range probed by their lensing maps.  Separating the linear bias from the redshift distribution, therefore, requires adopting a model for $df/dz$. Ref.~\cite{Thakore2025} adopts a redshift distribution whose best-fit model peaks at $z > 2$, with relatively little flux at the redshifts most strongly probed by DES lensing ($z \lesssim 0.5$). Under such an assumption, a larger effective bias is required to reproduce a given measured value of $b\, df/dz$. Consequently, while our tomographic analysis favors a lower inferred bias for the UGRB sources over $0 < z \lesssim 2$, it remains unclear whether this reflects a difference in the underlying cross-correlation amplitudes at $z < 1$, or instead arises from differing assumptions about the redshift distribution of the unresolved emission.

Relatedly, recent work cross-correlating the Fermi UGRB with KiDS lensing did not detect any significant cross-correlation \cite{Zhang2026}, but they do not appear to report upper limits on the bias of the UGRB sources, so it is unclear whether their results are consistent with our results.  At a qualitative level, however, we do agree that there are only weak correlations between the Fermi UGRB and cosmological large-scale structure at $z\lesssim 1$.

If we assume that most of the UGRB sources overlap in redshift with our galaxy samples (i.e., $f_{0-2.1}\approx 1$), then the inferred bias of the UGRB is quite low.  A glaring example is the 1-2 GeV bin, for which $b\approx 0.4$ is inferred, too small to be reconciled with the expected bias of any plausible host halo population.  Therefore we argue that the surprisingly low amplitude of the UGRB cross-correlation requires $f_{0-2.1} < 1$, which suggests the presence of emission at redshifts not covered by our DESI and unWISE samples at $0<z\lesssim 2$.  Two possible examples of such emission are residual contamination from gamma-ray sources within our own Milky Way galaxy, or an unknown population of high-redshift ($z > 2)$ gamma-ray sources.  The former scenario is supported by the significant Galactic contamination apparent in UGRB maps and also seen in the UGRB auto-correlation, and a crude estimate of the amplitude of Galactic contamination gives inferred bias values $b\approx 2$ for the bins at 2-10 GeV, consistent with the mean bias of the point source population.  However, we cannot exclude the latter scenario, namely that there is also significant $\gamma$-ray emission from sources at $z>2$, beyond the redshift range probed by our galaxy samples.  This scenario could be tested using cross-correlations with tracers of large-scale structure at $z>2$, and we suggest that CMB lensing might provide a suitable tracer.

\section*{Data Availability}
Data used in this paper
is available on Zenodo at
\url{https://zenodo.org/uploads/20536491}.

\acknowledgments

We thank Aurelio Amerio, Yi-Kuan Chiang, Alessandro Cuoco, Richard Feder, Nicolao Fornengo, Joshua Kim, Mathew Madhavacheril, Michela Negro, Anthony Pullen, and Marco Regis for helpful discussions.  This work was completed at the Kavli Institute for Theoretical Physics, and we thank all the participants of the KITP program ``Exploring New Boundaries in Cosmology and Astrophysics'' for encouragement.  
Research at Perimeter Institute is supported in part by the Government of Canada through the Department of Innovation, Science and Economic Development Canada, and by the Province of Ontario through the Ministry of Colleges and Universities.
AK was supported as a CITA Na-
tional Fellow by the Natural Sciences and Engineering
Research Council of Canada (NSERC), funding reference
\#DIS-2022-568580.
WP acknowledges support from the Natural Sciences and Engineering Research Council of Canada (NSERC), [funding reference number RGPIN-2025-03931] and from the Canadian Space Agency. 
This research was enabled in part by support provided by Compute Ontario (computeontario.ca) and the Digital Research Alliance of Canada (alliancecan.ca).

This material is based upon work supported by the U.S. Department of Energy (DOE), Office of Science, Office of High-Energy Physics, under Contract No. DE–AC02–05CH11231, and by the National Energy Research Scientific Computing Center, a DOE Office of Science User Facility under the same contract. Additional support for DESI was provided by the U.S. National Science Foundation (NSF), Division of Astronomical Sciences under Contract No. AST-0950945 to the NSF’s National Optical-Infrared Astronomy Research Laboratory; the Science and Technology Facilities Council of the United Kingdom; the Gordon and Betty Moore Foundation; the Heising-Simons Foundation; the French Alternative Energies and Atomic Energy Commission (CEA); the Secretariat of Science, Humanities, Technology and Innovation (SECIHTI) of Mexico; the Ministry of Science, Innovation and Universities of Spain (MICIU/AEI/10.13039/501100011033), and by the DESI Member Institutions: \url{https://www.desi.lbl.gov/collaborating-institutions}. Any opinions, findings, and conclusions or recommendations expressed in this material are those of the author(s) and do not necessarily reflect the views of the U. S. National Science Foundation, the U. S. Department of Energy, or any of the listed funding agencies.

The authors are honored to be permitted to conduct scientific research on I'oligam Du'ag (Kitt Peak), a mountain with particular significance to the Tohono O’odham Nation.

\bibliography{refs,DESI_supporting_papers}

\appendix
\section{Power spectra of the unresolved gamma-ray background}
\label{sec:corr_plots}

\begin{figure}
    \includegraphics[width=0.98\linewidth]{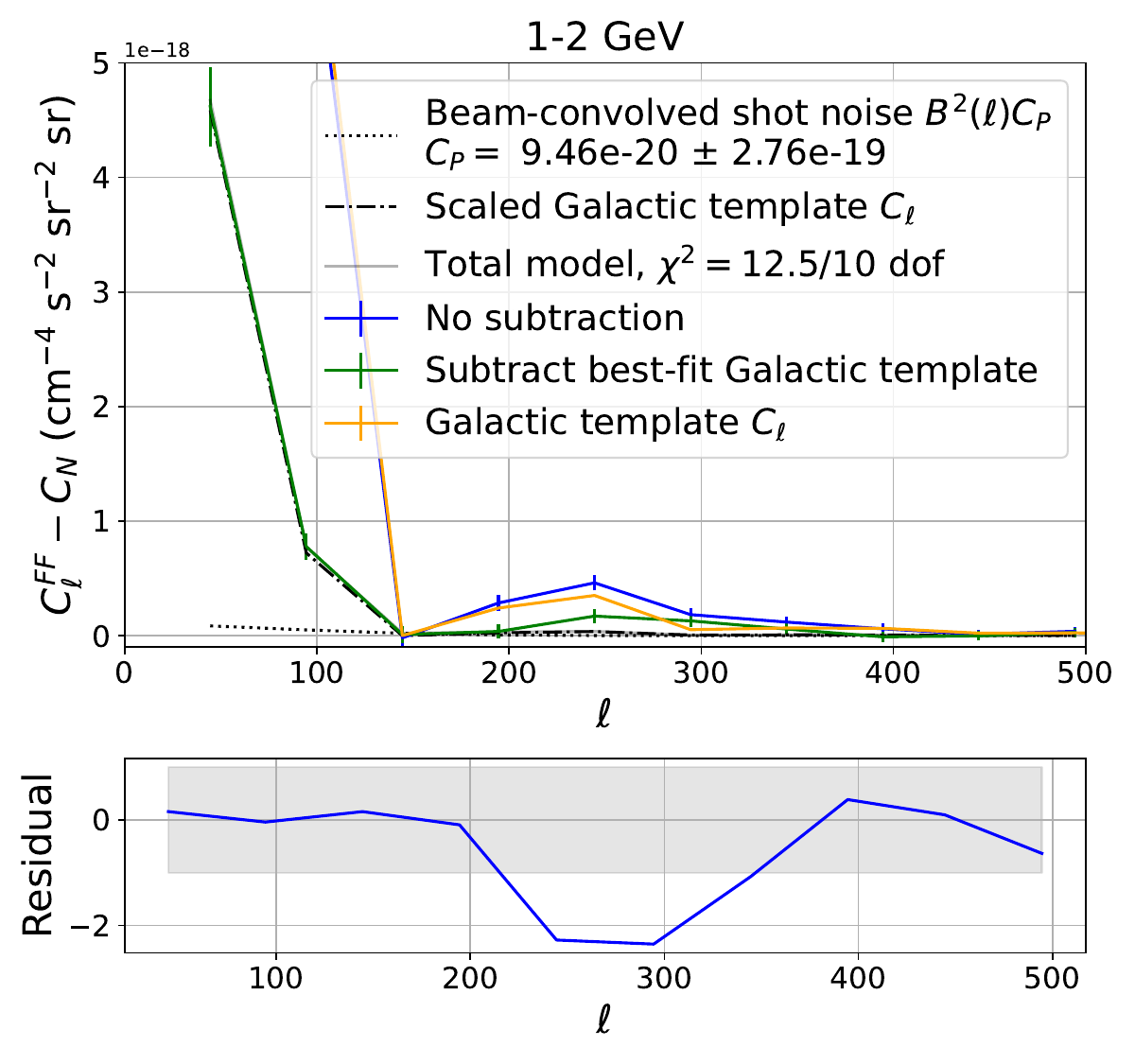}
    \caption{Auto-spectrum for UGRB energy bin 1-2 GeV.  The blue curve shows the autospectrum after subtracting the Poisson spectrum $C_N$ for the observed number of photons.  The green curve shows the auto-spectrum of the UGRB after subtracting a diffuse Galactic template \emph{at the map level}.  The orange curve shows the auto-spectrum of this subtracted template.  As discussed in the text, the large residual low-$\ell$ power in the green curve indicates that this subtraction does not remove all Galactic emission.  We estimate the amplitude of the residual Galactic contamination by fitting the green curve as a sum of a scaled version of the orange curve and scaled version of the square of the beam, shown in the black dash-dot and black dotted curves respectively.  The total model for $C_\ell$ (sum of two black curves) is shown in the grey curve, and the bottom panel shows the residuals between this total model and the green curve, scaled by the uncertainties. We use the normalizations of the two black curves to estimate the fraction of the UGRB map arising from residual Galactic contamination and from extragalactic emission.}
    \label{fig:Clff1}
\end{figure}

\begin{figure}
    \includegraphics[width=0.98\linewidth]{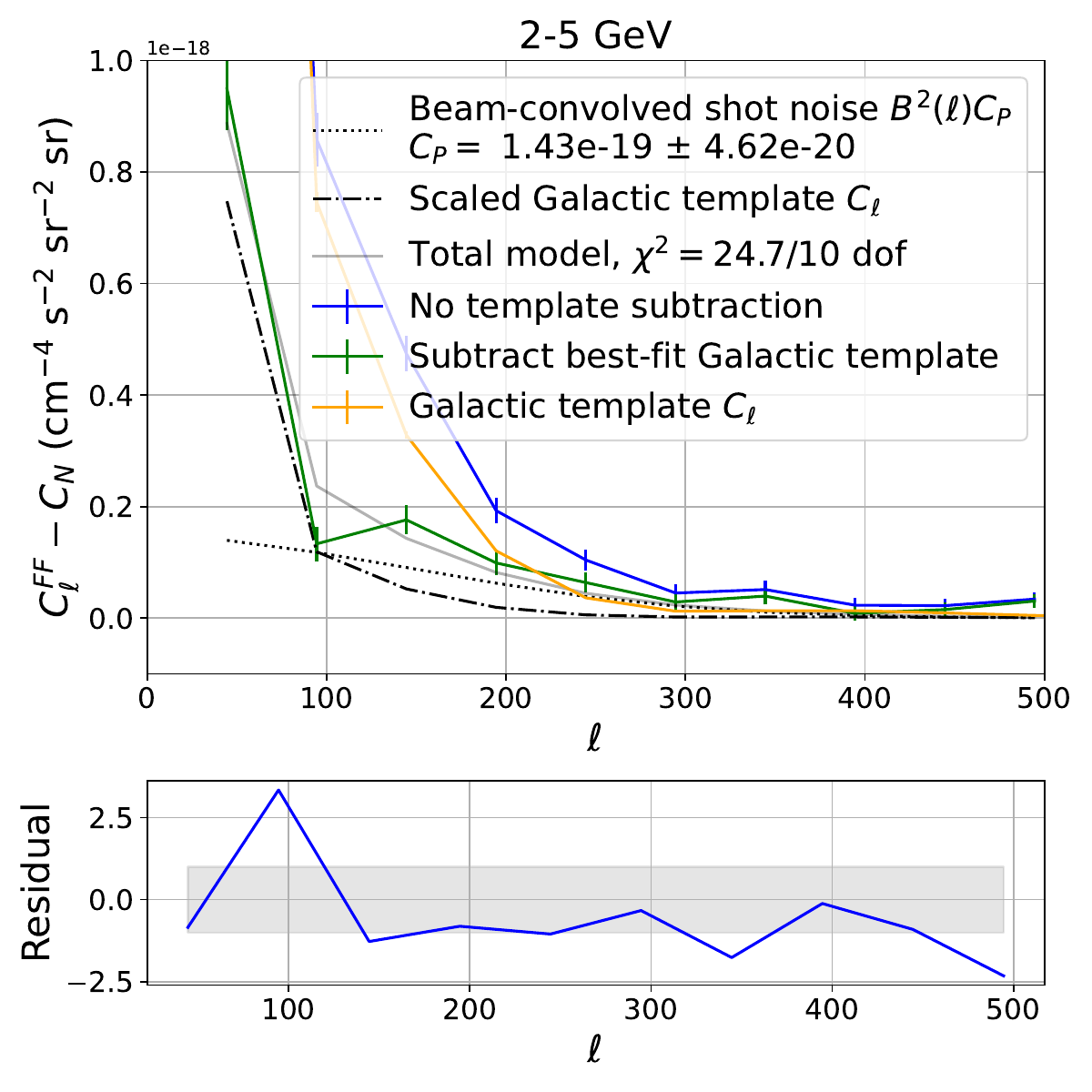}
    \caption{Similar to \cref{fig:Clff1}, but for the 2-5 GeV bin.}
    \label{fig:Clff2}
\end{figure}

\begin{figure}
    \includegraphics[width=0.98\linewidth]{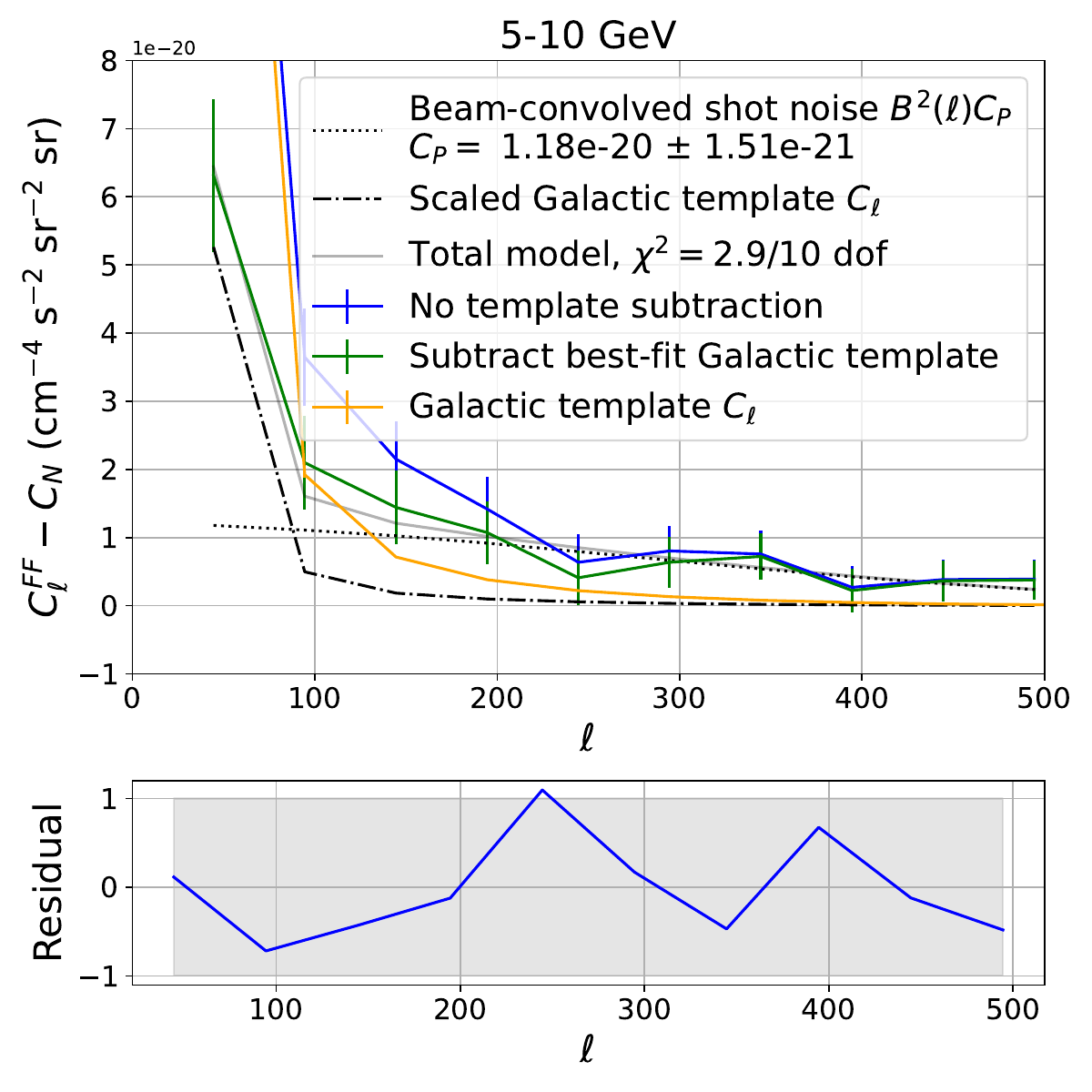}
    \caption{Similar to \cref{fig:Clff1}, but for the 5-10 GeV bin.}
    \label{fig:Clff3}
\end{figure}

\begin{figure}
    \includegraphics[width=0.98\linewidth]{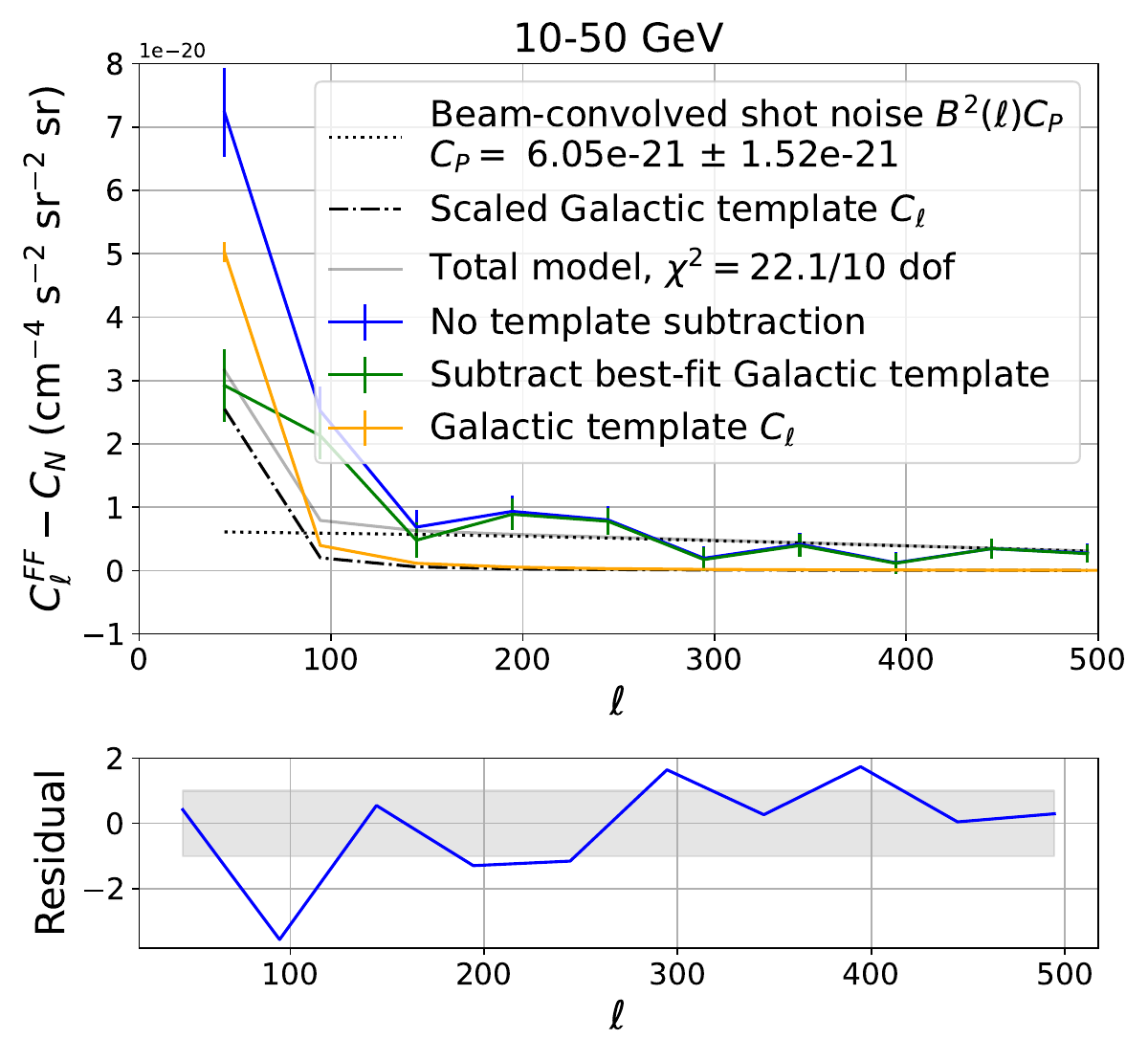}
    \caption{Similar to \cref{fig:Clff1}, but for the 10-50 GeV bin.}
    \label{fig:Clff4}
\end{figure}

\begin{figure*}
    \includegraphics[width=0.98\linewidth]{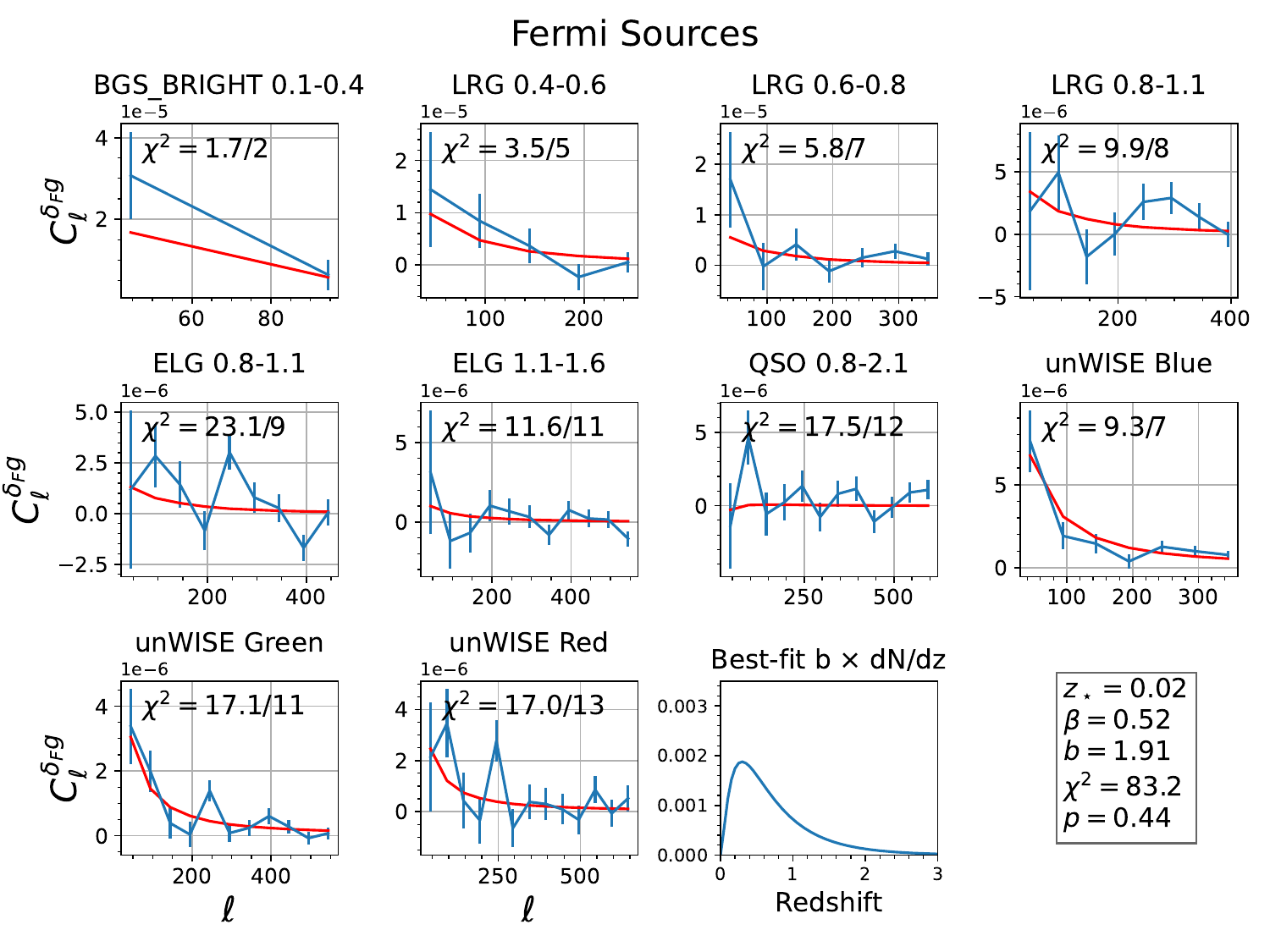}
    \caption{Best-fit model and data cross-spectra between the Fermi point sources and DESI and unWISE galaxies.}
    \label{fig:pointsourcecross}
\end{figure*}

\begin{figure*}
    \includegraphics[width=0.98\linewidth]{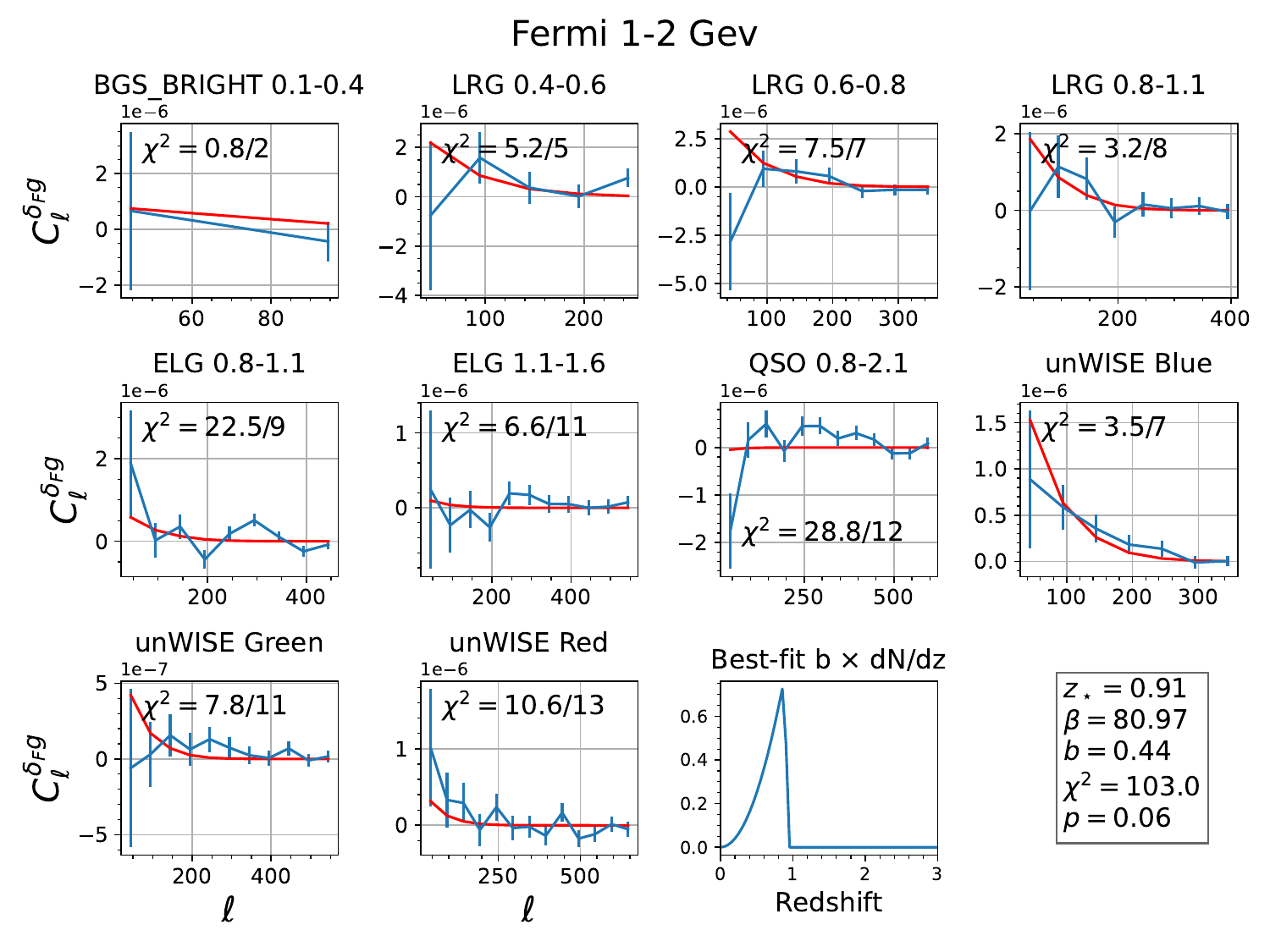}
    \caption{Measured cross-correlation between the Fermi diffuse map at 1 to 2 GeV and the unWISE and DESI galaxies (blue), and best-fit model (red). Text in each panel gives the $\chi^2$ of the model for that galaxy sample, and the number of data points used. The final panel shows the best-fit $b \times dN/dz$ and the text box to the right gives its parameters, overall $\chi^2$, and $p$-value.}
    \label{fig:bin_1_xcorr}
\end{figure*}

\begin{figure*}
    \includegraphics[width=0.98\linewidth]{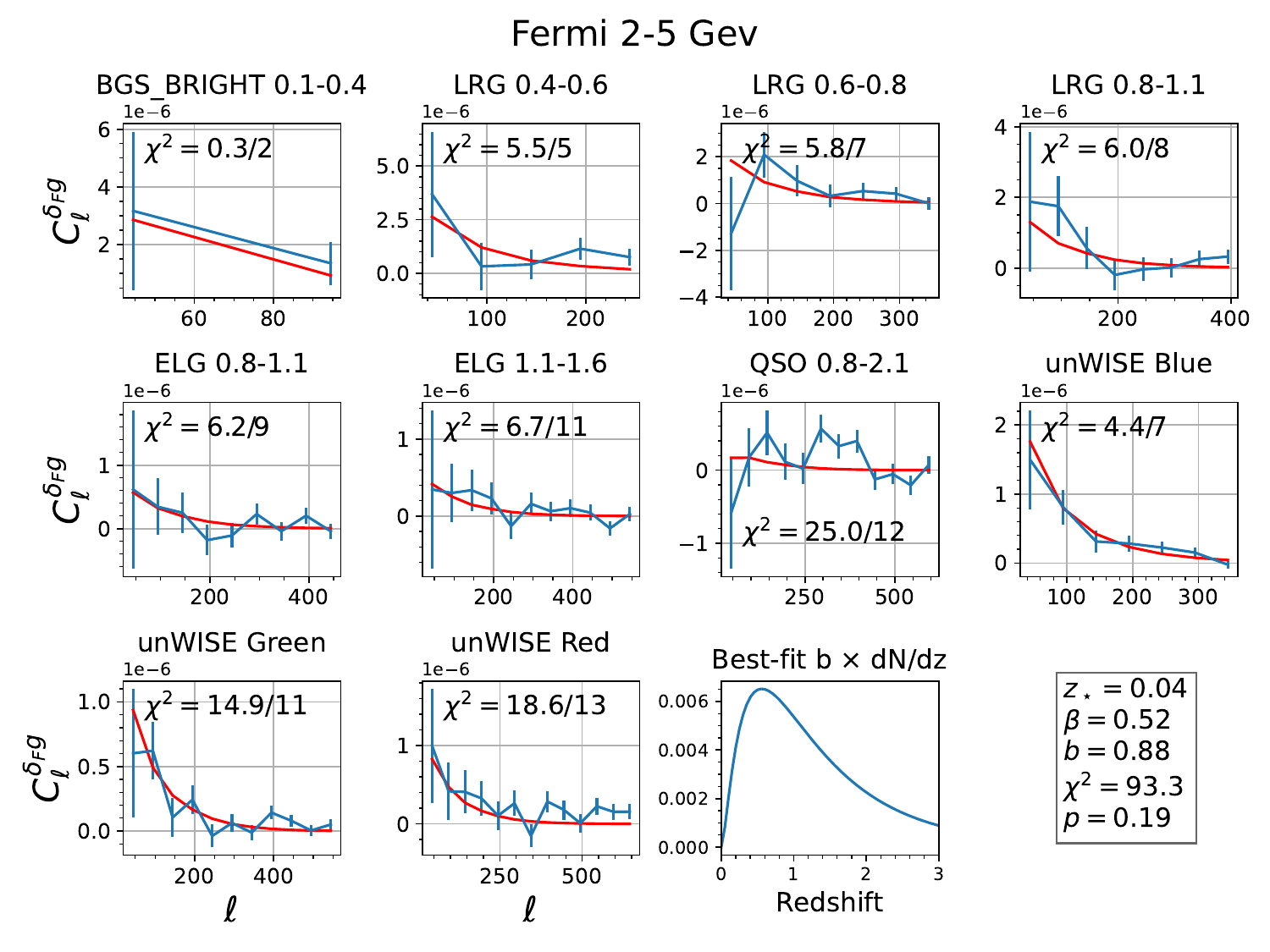}
    \caption{As in Fig.~\ref{fig:bin_1_xcorr}, but for the Fermi map at 2 to 5 GeV.}
    \label{fig:bin_2_xcorr}
\end{figure*}

\begin{figure*}
    \includegraphics[width=0.98\linewidth]{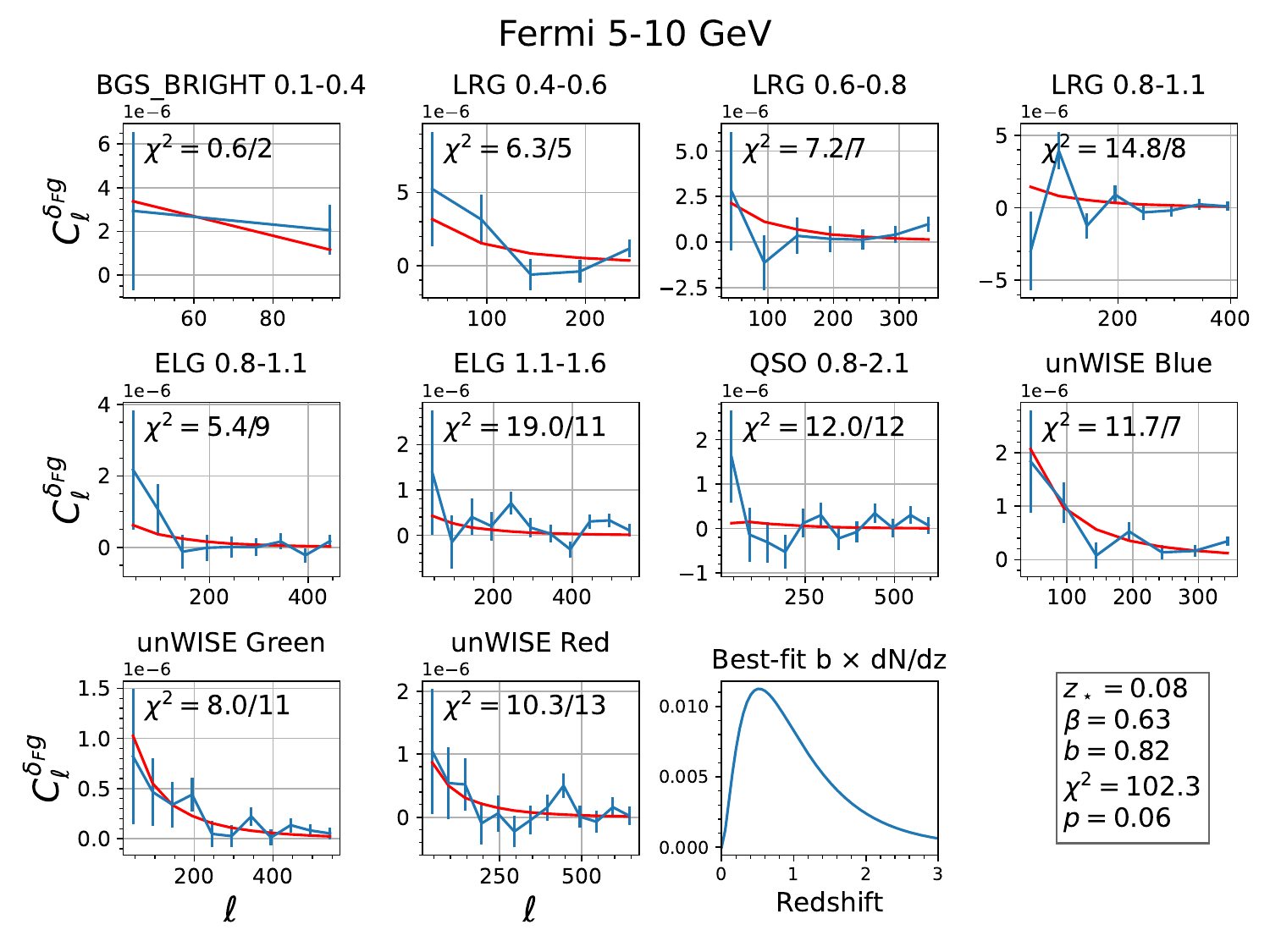}
    \caption{As in Fig.~\ref{fig:bin_1_xcorr}, but for the Fermi map at 5 to 10 GeV.}
    \label{fig:bin_3_xcorr}
\end{figure*}

\begin{figure*}
    \includegraphics[width=0.98\linewidth]{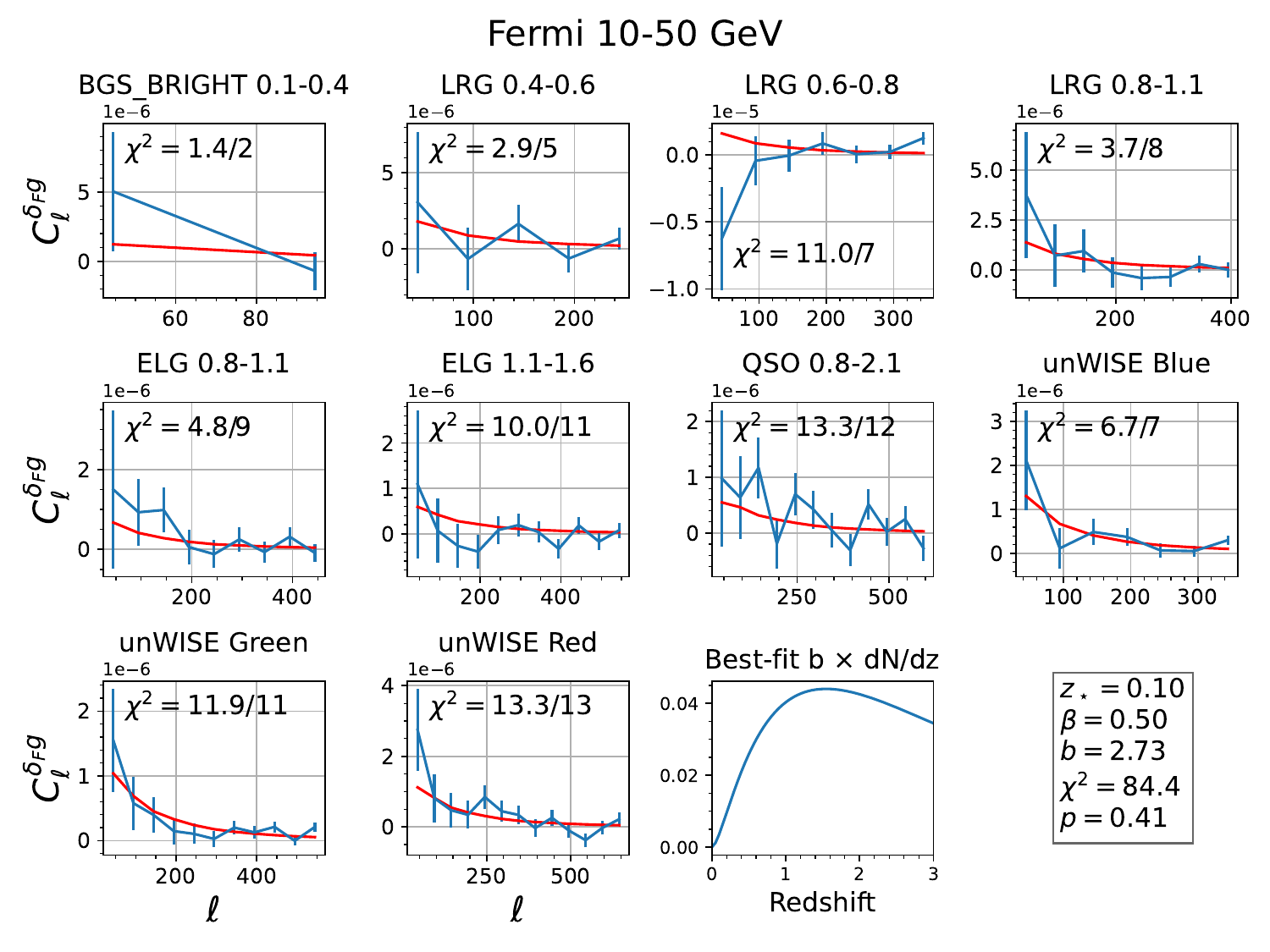}
    \caption{As in Fig.~\ref{fig:bin_1_xcorr}, but for the Fermi map at 10 to 50 GeV.}
    \label{fig:bin_4_xcorr}
\end{figure*}

In this appendix, we show plots of the auto-spectra of our Fermi $\gamma$-ray maps, as well as their cross-spectra with DESI and unWISE galaxies.  \cref{fig:Clff1,fig:Clff2,fig:Clff3,fig:Clff4} show the auto-spectra of four UGRB energy bins in the range 1-50 GeV.  As discussed in \cref{sec:autocorr} and below in \cref{sec:contamination}, these auto-spectra are used to estimate the fraction of the UGRB arising from residual Galactic contamination and the fraction from extragalactic emission. 
\cref{fig:pointsourcecross} shows the cross-spectra of the Fermi point source map with each of the large-scale structure (LSS) maps, while \cref{fig:bin_1_xcorr,fig:bin_2_xcorr,fig:bin_3_xcorr,fig:bin_4_xcorr} show similar cross-spectra for the same four UGRB energy bins. 
These cross-power spectra are used in \cref{sec:pointsrc,sec:ugrbcross} to derive population properties of the point sources and UGRB, respectively.

The quasar cross-spectra with the point sources and the 1--2 GeV bin are slightly negative; this is because the best-fit Fermi $b\,dN/dz$ is largely at lower redshifts than the quasar sample, so the cross-correlation is dominated by the magnification of the quasars, which yields a negative signal because the faint end slope is $<0.4$ for the quasars. While the overall $\chi^2$ are good for each of the five samples, we also note that, out of the 50 cross-spectra shown, three of them have marginally bad fits ($p < 0.01$), which is unlikely to occur by chance. Two of these failures are in the ELGs at $0.8 < z < 1.1$, which is the most susceptible DESI sample to imaging systematic contamination \cite{KP3s2-Rosado}, while the third failure is driven by the anomalously low $\ell < 70$ correlation between the 1--2 GeV bin and the DESI quasars. These failures may be indicative of residual Galactic contamination in the Fermi maps.

\section{Estimation of contamination fraction}
\label{sec:contamination}

\begin{table*}[!ht]
\centering
\begin{tabular}{l | c c c}
\hline
Energy bin & Contamination fraction &~~ $C_N$ (cm$^{-4}$ s$^{-2}$ sr$^{-2}$ sr) ~~ & ~~$C_P$ (cm$^{-4}$ s$^{-2}$ sr$^{-2}$ sr) ~~\\
\hline
1-2 GeV & $0.55 \pm 0.03$ & $2.7\times10^{-18}$ & $0.95 \pm 2.8\times10^{-19}$ \\
2-5 GeV & $0.63 \pm 0.06$ & $1.2 \times 10^{-18}$ & $1.4 \pm 0.46 \times10^{-19}$ \\
5-10 GeV & $0.58 \pm 0.04$  & $2.9 \times 10^{-19}$ & $1.2 \pm 0.15 \times10^{-20}$ \\
10-50 GeV & $0.57 \pm 0.11$ & $1.6 \times 10^{-19}$ & $6.05 \pm 1.5 \times 10^{-21}$ \\
\hline
\end{tabular}
\caption{UGRB auto-correlation parameters.  The contamination fraction is estimated from the amplitude of the non-Poissonian term, see text in \cref{sec:contamination} for more detail. The uncertainties are purely statistical and do not account for systematic errors arising from assumptions in this ad-hoc procedure, which likely dominate the statistical error.}

\label{tab:auto_results}
\end{table*}

To get a rough sense for what fraction of our unresolved emission map arises from residual Galactic contamination, we fit the map auto-spectrum to a sum of two terms: a shot noise power spectrum, and a power spectrum whose shape matches that of the auto-spectrum of the diffuse Galactic template that we subtracted.  We assume that the amplitude of the Galactic contamination term scales with the (square of) the amplitude of the residual Galactic contamination.  We stress that this is a completely ad hoc assumption, with no real justification, and we make use of this only to get an order-of-magnitude estimate for the fraction of the unresolved flux that arises from residual unsubtracted Galactic emission.

The shot noise term in our fit to the auto-spectrum also requires some discussion, because the form of this spectrum is not expected to be completely white, but instead should change in amplitude on scales of order the size of the point spread function (PSF).  On scales smaller than the PSF size, the shot noise amplitude is simply given by the mean number density of photons, but on scales larger than the PSF size, the number of sources also influences the shot noise. If the angular number density of sources that provide $N$ photons is $n(N)$, then on large scales, $\ell\theta\ll 1$ (where $\theta$ is the size of the PSF), the white noise amplitude is
\begin{equation} \label{eq:lowl}
C_\ell \approx \sum_{N=1}^{N_{\rm max}} N^2 n(N),    
\end{equation}
where $N_{\rm max}$ is the maximum number of photons that falls below the point source detection threshold.  In contrast, on small scales ($\ell\theta\gg1$), the white noise amplitude is
\begin{equation} \label{eq:highl}
C_\ell \approx \sum_{N=1}^{N_{\rm max}} N\, n(N).    
\end{equation}
Therefore, we expect the shot noise power spectrum to have a shape following the (square of the) shape of the PSF, multiplied by \cref{eq:lowl}, asymptoting to \cref{eq:highl} on scales smaller than the PSF size.  These are the expected white noise amplitudes for counts; these expressions are modified for flux maps by dividing by the square of the exposure map before taking the expected value of counts.

The ratio of \cref{eq:lowl,eq:highl} depends on the luminosity function of the sub-threshold sources.  As a concrete example, suppose that the number density of sources with mean expected number of photons $\mu$ is $n(\mu)$.  Since the actual number of photons $N$ is a random draw from a Poisson distribution with mean $\mu$, then we have
\begin{equation}
n(N) = \int d\mu \frac{dn}{d\mu} P(N|\mu) = 
\int d\mu \frac{dn}{d\mu} \frac{\mu^N}{N!}e^{-\mu}.
\end{equation}
Next let us write $dn/d\mu \propto \mu^{-\alpha}$, where $\alpha$ is the slope of the luminosity function.  Allowing for the point source detection threshold to vary over the sky, i.e.\ writing $N_{\rm max}$ as a function of sky coordinates $\Omega$, then the ratio of \cref{eq:lowl} to \cref{eq:highl} becomes
\begin{equation}
\frac{C_\ell(\ell\theta\ll 1)}{C_\ell(\ell\theta\gg 1)} \approx
\frac{\int d\Omega\sum_{N=1}^{N_{\rm max}(\Omega)} \frac{N^2}{N!}\Gamma(1+N-\alpha)}{\int d\Omega\sum_{N=1}^{N_{\rm max}(\Omega)} \frac{N}{N!}\Gamma(1+N-\alpha)}.
\end{equation}
Using a map of the point source detection threshold provided by \texttt{fermitools}, and extrapolating the slope of the luminosity function of detected point sources, we can predict the ratio of the two amplitudes \cref{eq:lowl,eq:highl}, thereby leaving only the overall normalization as the sole free parameter describing the shot noise spectrum.  Alternatively, if we can measure the two amplitudes from the observed auto-spectrum, this constrains the faint-end slope of the luminosity function.  We adopt the latter approach, allowing the two amplitudes of the shot noise spectrum to be free parameters.  We define $C_N$ as the shot noise amplitude at $\ell\,\theta_{\rm PSF}\gg 1$, and $C_P+C_N$ as the amplitude at $\ell\,\theta_{\rm PSF}\ll 1$. $C_P$ is the shot noise arising from fluctuations in the number of sources, defined in the same way as \cite{Ackermann2018}.  We therefore fit the auto-spectra of the UGRB maps (after subtracting the best-fit MW template) using a model with 3 parameters, 
\begin{equation}    \label{eq:auto-model}
C_\ell = \left(a \, C_{\rm MW}+ C_P\right) W_\ell^2 + C_N ,
\end{equation}
where $W_\ell$ is the PSF and $C_{\rm MW}$ is the power spectrum of the Milky Way diffuse emission template.  We then estimate the fraction of $\gamma$-ray flux arising from residual Galactic contamination as $f_{\rm cont} \approx a^{1/2} F_{\rm MW}/F_{\rm tot}$, where $F_{\rm MW}$ is the total flux from the subtracted Galactic template, and $F_{\rm tot}$ is the total flux in the UGRB map after subtracting the Galactic template.  The resulting fitted parameters are given in \cref{tab:auto_results}.  In general, we find significant residual contamination remaining after subtraction of Galactic foreground templates, with estimated fractions typically $f_{\rm cont} \approx 60\%$. We also find $C_P\ll C_N$, suggesting that the unresolved UGRB largely arises from sources each emitting $\sim 1$ photon, consistent with the steep slope ($\approx -1.92$) of the luminosity function found for resolved sources \cite{Daylan2017,Amerio2024}.

Even after subtracting the best-fit template, Fig.~\ref{fig:fermi_maps_galactic_subtracted} still shows clear structure correlated with the Galaxy near the Galactic center, which is most evident in the highest-energy bins. We therefore tested a variant mask where we also exclude  $\ell < 60^\circ$ or $\ell > 280^\circ$, thus removing these Galactic residuals and reducing $f_{\textrm{sky}}$ for the Fermi maps by $\sim30\%$. When fitting the Fermi auto-power spectrum measured in this mask, we find reduced contamination fractions of $\sim$35-40\%, implying that these residuals are responsible for some of the Galactic contamination in the unresolved background, but there is still substantial Galactic contamination further from the Galactic center.

These fits to the auto-spectrum are shown in \cref{fig:Clff1,fig:Clff2,fig:Clff3,fig:Clff4}.
The low-$\ell$ excess power is broadly consistent with previous work, e.g.\ Fig.~7 in \cite{Ackermann2018}, although we generally find excess power above the shot-noise components at higher $\ell$ than in \cite{Ackermann2018}, perhaps because our analysis makes use of updated Fermi maps and source masking that were not yet available in that earlier analysis.  Our fits indicate evidence for Galactic contamination even at $\ell > 50$, as seen in \cref{fig:Clff1,fig:Clff2}. In our fits, the power spectra from the Galactic template and from white noise are somewhat degenerate at low energy, where the PSF is large, because both are declining functions of $\ell$ even on relatively large scales. By allowing for nonzero contamination, our fits favor lower values of $C_P$ than \cite{Ackermann2018}  in the 1--2 GeV and 2--5 GeV bins, where the Galactic contamination power spectrum is largest. We find good consistency with \cite{Ackermann2018} at $>$5 GeV. This low-energy contamination would weaken the preference for a double-power law fit to the anisotropy energy spectrum as compared to a single power law.

\end{document}